\documentclass[aps,prb,showpacs,superscriptaddress,twocolumn,amsmath]{revtex4}
\usepackage{txfonts}
\usepackage{graphicx}
\usepackage{bm}
\usepackage{overpic}
\usepackage{bbold}
\def\rnum#1{\expandafter{\romannumeral #1}} 
\def\Rnum#1{\uppercase\expandafter{\romannumeral #1}}

\newfont{\bg}{cmr10 scaled\magstep4}

\newcommand{\bigzerou}{\smash{\lower1.8ex\hbox{\bg 0}}}

\begin{document}

\title{Order, disorder and tunable gaps in the spectrum of Andreev bound states in a 
multi-terminal superconducting device}

\author{Tomohiro Yokoyama}
\email[E-mail me at: ]{tomohiro.yokoyama@issp.u-tokyo.ac.jp}
\affiliation{Kavli Institute of Nanoscience, Delft University of Technology,
Lorentzweg 1, 2628 CJ, Delft, The Netherlands}
\affiliation{The Institute for Solid State Physics, The University of Tokyo,
5-1-5, Kashiwa-no-ha, Kashiwa, Chiba, 277-0882, Japan}

\author{Johannes Reutlinger}
\affiliation{Fachbereich Physik, Universit\"{a}t Konstanz, D-78457 Konstanz, Germany}

\author{Wolfgang Belzig}
\affiliation{Fachbereich Physik, Universit\"{a}t Konstanz, D-78457 Konstanz, Germany}

\author{Yuli V.\ Nazarov}
\affiliation{Kavli Institute of Nanoscience, Delft University of Technology,
Lorentzweg 1, 2628 CJ, Delft, The Netherlands}

\date{\today}

\begin{abstract}
We consider the spectrum of Andreev bound states (ABSs) in an exemplary 4-terminal superconducting structure
where 4 chaotic cavities are connected by quantum point contacts to the terminals and to each other forming a ring.
We nickname the resulting device {\it 4T-ring}.
Such a tunable device can be realized in a 2D electron gas-superconductor or a graphene-based hybrid structure.

We concentrate on the limit of a short structure and large conductance of the point contacts
where there are many ABS in the device forming a quasi-continuous spectrum.
The energies of the ABS can be tuned by changing the superconducting phases of the terminals.
We observe the opening and closing of gaps in the spectrum upon changing the phases.
This concerns the usual proximity gap that separates the levels from zero energy as well as
less usual ``smile'' gaps that split the levels of the quasi-continuous spectrum.

We demonstrate a remarkable crossover in the overall spectrum that occurs upon changing
the ratio of conductances of the inner and outer point contacts.
At big values of the ratio (closed limit), the levels exhibit a generic behavior expected for the spectrum of
a disordered system manifesting level repulsion and Brownian ``motion'' upon changing the phases.
At small values of the ratio (open limit), the levels are squeezed into narrow bunches separated by wide smile gaps.
Each bunch consists of almost degenerate ABS formed by Andreev reflection between two adjacent terminals.

We study in detail the properties of the spectrum in the limit of a small ratio,
paying special attention to the crossings of bunches.
We distinguish two types of crossings:
i. with a regular phase dependence of the levels and
ii. crossings where the Brownian motion of the levels leads to an apparently irregular phase-dependence.
We work out a perturbation theory that explains the observations
both at a detailed level of random scattering in the device and
at a phenomenological level of positively defined random matrices.

The unusual properties of the spectrum originate from rather unobvious topological effects.
Topology of the first kind is restricted to the semiclassical limit and
related to the winding of the semiclassical Green's function.
It is responsible for the closing of the proximity gaps.

Topology of the second kind comes about the discreteness of the number of modes in the point contacts
and is responsible for the smile gaps.
The topology of the third kind leads to the emergence of Weyl points in the spectrum
and is not discussed in the context of this article.

\end{abstract}
\maketitle

\section{INTRODUCTION}

Disordered and random systems play a fundamental role in a broad range of research fields.
Early works concentrated on the spectra of complex atomic nuclei, which
could be described by random Hamiltonians, and led to the development of random matrix theory (RMT)~\cite{dyson:62}.
Due to the quantum mechanical effect of level repulsion combined with universality, RMT leads
to the famous Wigner-Dyson distribution of level spacings~\cite{wigner:57}.
Remarkably, such level distributions depend only on the symmetries of the system.
The connection to solid state physics has been made in the context of localization~\cite{anderson:80},
small particles~\cite{larkin:75} and mesoscopic effects like weak localization or conductance fluctuations~\cite{altshuler:91}.
Interestingly, much less  works have addressed the spectral properties near a gap in the spectrum, which is especially relevant in superconducting systems~\cite{melsen:96,lodder:98}.
Here, the universal fluctuations have been predicted as well~\cite{tracy:94}.
These predictions  have been confirmed numerically for proximity systems~\cite{vavilov:01}.
So-called Andreev billiards have henceforth provided an important playground to test general chaotic dynamics~\cite{beenakker:05}.

The superconducting proximity effect in general describes the physical properties of normal,
non-superconducting conductors in close electronic contact to a superconductor.
In this way, an otherwise normal structure can show the key features of superconductivity,
such as perfect diamagnetism, a supercurrent or a spectral gap~\cite{nazarov:99,belzig:99}.
Particularly the induced gap and its phase-dependence has attracted attention
from theoretical~\cite{belzig:96,zhou:98} and experimental ~\cite{gueron:96,lesueur:08} side.
In the long junction limit $L\gg\xi_S$ with the superconducting coherence length $\xi_S$,
the gap scales in a metal of size $L$ is universally the Thouless energy $E_{Th}\sim D/L^2\sim \hbar/\tau_{d}$,
$D$ being the diffusion constant and $\tau_d$ being the dwell time inside the normal metal.
The induced gap is called the ``minigap'', since in the long junction regime it is usually much smaller than
the superconducting gap $\Delta$, though in practical realizations this is not always the case.
In particular Le Sueur and coworkers~\cite{lesueur:08} have measured the phase-dependent local density of states in
a diffusive wire between superconducting contacts and found an excellent agreement with
the theory based on quasiclassical Greens functions.
The size of the minigap depends in a characteristic way on the phase difference and closes for a phase difference of $\pi$.

On the microscopic level, the electronic connection between superconductor and a normal metal stupilates the process of Andreev reflection~\cite{andreev:64}, in which an electron-like quasiparticle is converted into
a Cooper pair leaving behind an hole-like quasiparticle.
This coherent processes can occur for energies below the superconducting gap $\Delta$ and
results in the presence of superconducting correlations in the normal metal.
For a finite-size normal metal or a junction between two superconductors subsequent Andreev processes form Andreev bound states (ABSs).
These bound states depend on the phase difference $\varphi$ between
the superconducting order parameters and, hence, carry a superconducting current.
This is the microscopic origin of the Josephson effects.
The discrete ABS have been detected by tunneling spectroscopy in carbon nanotube quantum dots~\cite{pillet:10}
and microwave spectroscopy in atomic point contacts~\cite{bretheau:13}.
These observations are in good agreement with theoretical predictions and
confirm the microscopic relation between phase-dependent Andreev states and the Josephson effect.

In view of the history of the superconducting proximity effect both in theory and experiment,
one would expect that everything is known already at least on the qualitative level.
Hence, it came as a complete surprise that Reutlinger \textit{et al.} have reported a secondary gap in the spectrum
just below the edge to the continuum in a short Josephson junction made from a chaotic cavity
connected to two superconductors~\cite{reutlinger:14a}.
Due to the peculiar phase dependence of the secondary gap, which closes for some critical phase difference,
it was termed smile gap and found to be parametrically related to the (small) factor $\Delta/E_{Th}$.
Furthermore, the presence or absence of the smile gap was related to the distribution of the transmission probabilities of
the two contacts connecting the cavity to the two superconductors~\cite{reutlinger:14b}.
The smile gap is present if the transmission distribution of each contact is  gapped at small transmissions, that is, there is a lower boundary for transmission eigenvalues.
The fact that the smile gap is robust against distortions of the transmission distributions and/or
the formation of multiple cavities suggests a universal mechanism for the formation of the smile gap in
systems of cavities connected to superconductors.

More recently the focus of research has moved towards multi-terminal superconducting devices.
A particular path-breaking prediction by Riwar and coworkers~\cite{riwar:16} was
the realization of Weyl-type topological matter.
The Weyl singularities can be engineered artificially in systems consisting of a quantum coherent conductor connected to
at least four superconducting terminals.
It is interesting to note that three terminals are insufficient to create topological points,
but still provide interesting physics~\cite{Heck,padurariu:15}.
The potential of engineering Weyl singularities in multi-terminal
Josephson junctions~\cite{riwar:16,YokoyamaNazarov2015}
is still at its infancy and many interesting possibilities need to be explored.
Thus, the fundamental properties of these systems need to be investigated, which we will address in this paper.

Many properties of Josephson junctions rely on the presence and properties of ABSs,
which are phase-dependent and current-carrying states connected to at least two superconducting terminals.
Such states are described by a surprisingly simple formula, derived by Beenakker~\cite{beenakker:97}.
The ABSs are represented by scattering matrices of electrons and holes propagating through
the non-superconducting part of the junction.
The transport characteristics of this normal region determines the properties of the ABSs and the Josephson current.
Andreev reflection at the superconductors, which converts electrons into holes and vice versa, can also be
expressed in terms of scattering matrices.
The combination of all scattering matrices results in an eigenvalue problem for the energies of ABSs, which is known as Beenakker's formula~\cite{beenakker:97}.
It is straightforwardly extended to multi-terminal junctions for our purposes.

In this paper, we study a particular multi-terminal Josephson junction based on a ring structure formed by connecting four chaotic cavities with a big number of channels to each other and to four superconducting terminals. We nickname it a 4T-ring. Despite the fact that the setup looks rather specific, we argue that the device illustrates interesting and general properties of the ABS in multi-terminal junctions. The 4T-ring can be experimentally realized in either 2D gas semiconducting heterostructures or in graphene.

We approach the ABS spectrum in two complementary ways.
One way is to use semiclassical Green's functions and quantum circuit theory~\cite{NazarovBlanter},
which leads to a continuous density of ABS energies below the superconducting gap $|\epsilon| \leq \Delta$ rather than a discrete spectrum.
We focus on the density of states at $\epsilon =0$ and reveal the presence of proximity gaps and gapless regions in the spectrum, as well as smile gaps.
The Green's function approach allows for the definition of topological numbers, which distinguish the gapped and gapless regions in
the 3D space of superconducting phases.
A complementary formulation is Beenakker's determinat equation~\cite{beenakker:97} for a random scattering matrix with a large number of channels,
allowing for the evaluation of individual ABS energies.
It explicitly demonstrates the gapped structure of the Andreev spectrum. Both the proximity and smile gaps are consistent with quantum circuit theory calculationes. We reveal the topology based on the gaps in the transmission distributions of individual cavities and explain the smile gaps with it.
In addition, the scattering matrix approach enables us to artificially break the gap in the transmission distribution by injecting a single transmission eigenvalue into this gap.
We demonstrate that this results in stray levels within the smile gaps.

We observe that the Andreev spectrum in our system is crucially tuned by the ratio of conductances, or
conduction channels, in the ring structure and in the contacts attaching to the superconductors.
If the conductances to the terminals are much larger than those in the ring (open limit),
the spectrum forms bunches with a finite number of Andreev levels, equal to the number of channels in the internal connector.
The bunches are localized between adjacent terminals and follow the phase differences of the involved superconductors.
The crossings between bunches are classified into two types, their properties can be understood from a perturbative treatment for the degenerate levels.
We investigate the visibility of wiggle-like fluctuations in the ABS phase dependence that stems from the general predictions of RMT for our disordered system.

The structure of the article is as follows.
In Section \ref{sec:setup}, we describe the setup of the 4T-ring
using quantum circuit theory as well as the scattering matrix approach.
In Section \ref{sec:overview}, we give an overview of the Andreev spectrum based on both descriptions, explain the limiting cases, formulate the topologies and explain their applications in understanding the spectrum.
Section \ref{sec:openlimit} is devoted to a specific discussion of the spectral details in the open limit, where we develop and apply a perturbation theory for the crossings of level bunches and investigate the fine structure of the bunches.
We conclude in Section \ref{sec:conclusions}.

\section{The 4T-ring}
\label{sec:setup}

In this Section, we describe the design of the 4T-ring,
and its description in the languages of quantum circuit theory and scattering matrices.

\subsection{Design}

\begin{figure}
\includegraphics[width=\columnwidth]{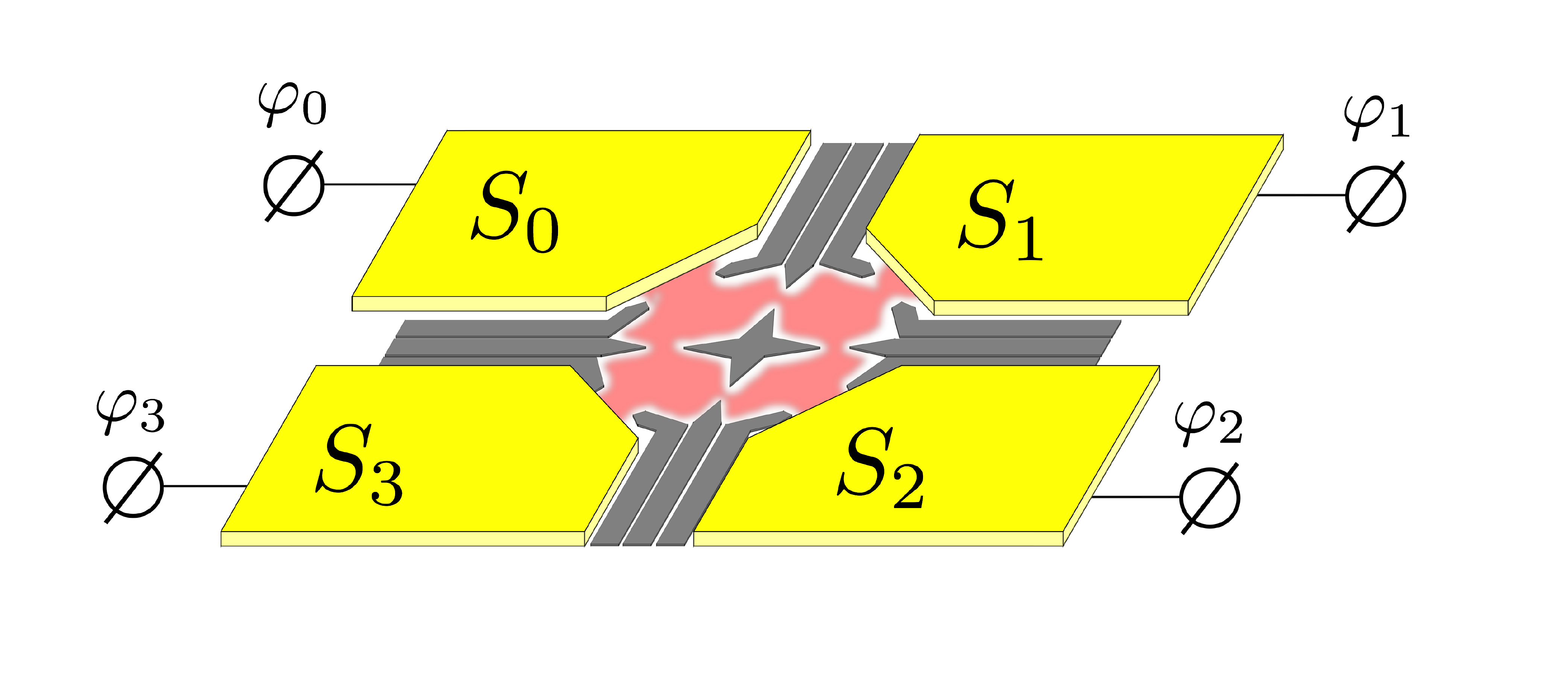}
\caption{(Color online)
Sketch of a possible experimental realization of a 4T-ring geometry using a 2D electron gas.
Multiple gates (gray regions) are used to deplete the 2D gas and to form four chaotic cavities,
which are connected to each other and to the superconducting terminals (yellow regions) through ballistic contacts.
$\varphi_i$ ($i=0,1,2,3$) indicates the superconducting phase in the terminal $i$.
A challenge might be the realization of the central gate,
that has to be contacted without disturbing the other connections.
}
\label{fig:Design}
\end{figure}

The nano-device we propose and discuss throughout the article is a hybrid superconducting-normal metal heterostructure.
There are four independent superconducting leads coming to the structure,
which serve as superconducting terminals, numbered with $k=0,1,2,3$.
The normal metal part consists of four chaotic cavities.
Each cavity is connected with a corresponding terminal by a ballistic contact
encompassing $N_i$ transport channels.
In addition, the cavities are connected to each other by ballistic contacts to form a ring-structure.
The number of transport channels in the contact between the cavities $i$ and $i+1$ is  $M_i$.
$i+1 =4$ denotes cavity $0$. In the following, we call this device a 4T-ring.
A sketch of the setup is shown in Fig.\ \ref{fig:Design}.

The design of the device is robust against inevitable imperfections of fabrication.
There is an additional contact resistance between the normal part and the superconductors:
yet it can be disregarded provided it is much smaller than the resistance of the ballistic contact.
The cavities are assumed to be fully chaotic and thus described by scattering matrices taken from
the circular ensemble of random matrix theory (RMT)~\cite{brouwer:95}.
The origin of chaoticity can be either due to impurity scattering inside the cavity or
due to scattering at the boundaries in an otherwise ballistic system with a rather arbitrary shape.
In both cases the resistance of the cavity interior must be much smaller than the resistance of the point pontacts.

We assume the short structure limit, that is, the electron dwell time inside the structure is shorter than $\hbar/\Delta$,
$\Delta$ being the superconducting energy gap in the leads.
This is known to be equivalent to the assumption of no energy-dependence of the scattering matrix of
the structure at a scale of $\Delta$,
which permits efficient numerical calculations of the energy spectrum of excitations in the structure.
We assume the superconducting leads made of the same metal.
In this case, the superconducting order parameters in the terminals $\Delta_i$ have
the same absolute value $\vert \Delta \vert$, but in general different phases $\varphi_i$.
Since physical effects depend only on phase-differences, one phase can be chosen to $0$,
which gives three parameters governing the spectrum in the device.

Experimentally the device can be realized on the basis of a semiconductor heterostructure supporting
a 2-dimensional electron gas (2DEG) at its interface, for instance, on the basis of  GaAs/AlGaAs heterostructures.
In ballistic 2DEGs point contacts of ideal transmission have been realized~\cite{wees:88, hartog:98}
and furthermore these systems can be coupled to superconducting leads,
allowing for an investigation of the proximity effect~\cite{wees:96, nguyen:92,choi:05},
where the importance of sufficiently transparent interfaces between the superconductor
and the 2DEG has been outlined.

A sketch of an idea for a experimental realization using a 2DEG is shown in Fig.\ \ref{fig:Design}.
The big yellow regions are the superconducting leads, the red, blurry region sketches the 2DEG within
the normal part of the structure. Gates are used to form the cavities with point-contacts between
each other and towards the superconductors with a variable number of channels.
An experimental challenge might be the central gate, which must be contacted from the back of the sample
or an isolated electrode contacted by an air-bridge technique~\cite{Yacoby}.
By varying the gate voltages, the number of modes in each contact can be controlled separately.
The superconducting phase-differences of the leads can be controlled by
superconducting loop structures (not shown in the plot), where a magnetic flux can be used to
adjust the phase differences.

An alternative idea is to make a device on the basis of a graphene flake~\cite{Morpurgo}.
The geometry in this realization would be very similar to that shown in Fig.\ \ref{fig:Design}.

\subsection{Circuit-theory representation}
\label{subsection:circuitthy}

\begin{figure}
\includegraphics[width=0.8\columnwidth]{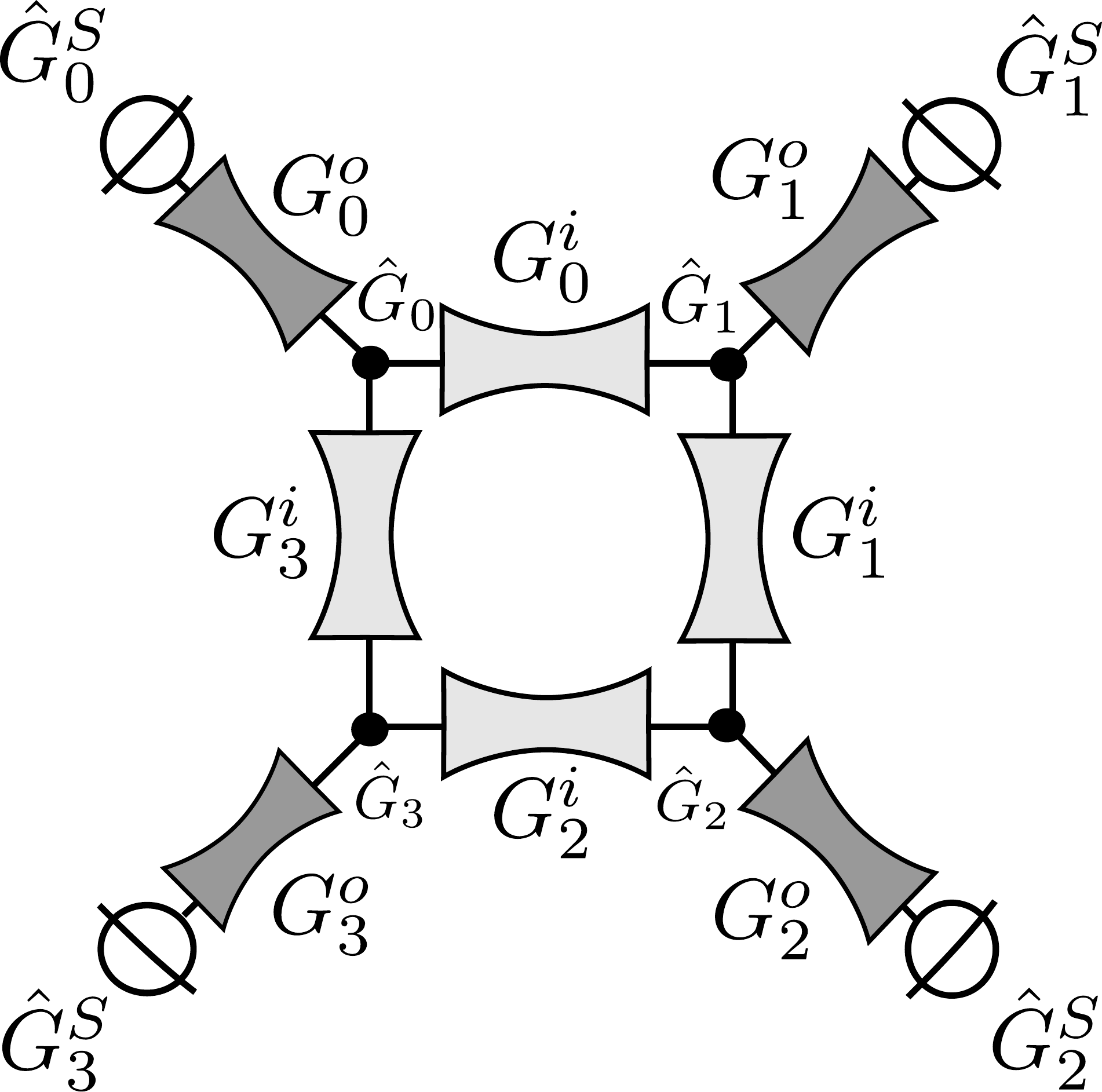}
\caption{
Circuit-theory scheme of the 4T-ring.
$G_k^{\rm i (o)}$ denotes the conductance of the inner (outer) connector $k$.
The nodes at the ends of the connectors and the superconducting terminals are characterized by
matrix voltages $\hat{G}_k$ and $\hat{G}_k^{\rm S}$, respectively.
}
\label{fig:ct-scheme}
\end{figure}

An elaborated unified description of quantum transport in nanostructures is provided by
quantum circuit theory~\cite{NazarovBlanter} that is valid in the semiclassical limit where
the typical conductance of the nanostructure by far exceeds the conductance quantum $G_{\rm Q}$.
In the circuit theory approach the nanostructure is separated into nodes, terminals, and connectors.
A matrix voltage (a matrix $\hat{G}$ satisfying $\hat{G}^2 = \hat{1}$,
${\rm Tr} \hat{G} =0$) is defined in each node and terminal of the structure.
The connectors are characterized by the distribution of transmission eigenvalues,
and the matrix currents in the connectors are expressed in terms of matrix voltages at
the ends of the connector. The matrix voltages in the terminals are fixed.
The matrix voltages in the nodes are found from matrix current conservation --- Kirchoff rules --- in the nodes.
These Kirchoff rules can be obtained from the extremization of an action ${\cal S}$ that
is contributed by each connector of the nanostructure and is a function of matrix voltages~\cite{NazarovBlanter}.

For the 4T-ring, four superconducting leads are regarded as four terminals, and the four cavities are treated as
four nodes (Fig.\ \ref{fig:ct-scheme}). Eight connectors involved are purely ballistic ones, with $T_p=1$.
The parameters of the device are the conductances of the connectors.
We denote matrix voltage in the leads by $\hat{G}^{\rm S}_k$, $k=0, \cdots, 3$,
and in the corresponding nodes just by $\hat{G}_k$. With this, the full action reads:
\begin{eqnarray}
\label{eq:action}
G_{\rm Q} {\cal S} = \sum_k G^{\rm o}_k s_{\rm B} (\hat{G}_k,\hat{G}^{\rm S}_{k})
+ \sum_k G^{\rm i}_k s_{\rm B} (\hat{G}_k,\hat{G}_{k+1})
\end{eqnarray}
where the ballistic connector action reads
\begin{equation}
s_{\rm B} (\hat{G}_A, \hat{G}_B) \equiv \frac{1}{2} {\rm Tr}
\{ \ln[ 1 + (\hat{G}_A \hat{G}_B + \hat{G}_B \hat{G}_A -2)/4 ] \}
\end{equation}
and $k+1=0$ for $k=3$.
The conductance is related to the number of channels in the corresponding contact,
$G^{\rm o}_k = G_Q N_k$,
$G^{\rm i}_k = G_Q M_k$.
It is enough for our purposes to keep the conductances of all outer and inner connectors
approximately the same, $G^{\rm i}_k \approx G^{\rm i}$, $G^{\rm o}_k \approx G^{\rm o}$.
The ratio of these two conductances, $G^{\rm i}/G^{\rm o}$ is an important parameter of our device,
its change influences the properties of the spectrum drastically.

To access the spectral properties of the ABS in the device,
it is enough to consider $2\times 2$ matrix voltages that are related to the energy-dependent semiclassical
advanced Green's function with Nambu indices. In the superconducting terminals
\begin{equation}
\hat{G}^{\rm S}_k(\epsilon) = \frac{1}{\sqrt{\Delta^2-(\epsilon + i0)^2 }}
\left[ \begin{array}{cc}
-i \epsilon & \Delta e^{i\varphi_k} \cr \Delta e^{-i\varphi_k} & i\epsilon
\end{array}\right],
\end{equation}
provided that the superconducting energy gap $\Delta$ is the same in all terminals.
In this case, the ABS energies are conveniently localized in the energy interval $0 < \epsilon < \Delta$.
There are three independent superconducting phases in the terminal, to fix the choice, we set $\varphi_0 = 0$.
In the nodes, the matrix voltage can be conveniently parametrized as
\begin{equation}
\label{eq:Gparametrization}
\hat{G}_k (\epsilon) =
\left[\begin{array}{cc}
\sin \theta_k & \cos \theta_k e^{i\eta_k} \cr \cos \theta_k e^{-i\eta_k} &-\sin \theta_k
\end{array}\right],
\end{equation}
where $\theta,\eta$ are real at $\epsilon = 0$. The local density of states in the node is given by
$\nu_0 {\rm Re}(\sin \theta_k)$, $\nu_0$ being the density of states in the normal metal.
To account for the presence of electronic states in the nodes, that leads to energy-dependent decoherence
between Andreev-reflections, one adds to the nodes so-called ``leakage'' terminals~\cite{NazarovBlanter}
that give extra terms in the action
\begin{equation}
{\cal S}_{\rm leak} = i \pi \epsilon \nu_0 \sum_k {\cal V}_k {\rm Tr} [\sigma_z \hat{G}_k]
\end{equation}
${\cal V}_k$ being the volume of node $k$.
We assume a ``short'' nanostructure where decoherence can be neglected,
and therefore neglect ${\cal S}_{\rm leak}$.
Comparing ${\cal S}$ and ${\cal S}_{\rm leak}$ at $\epsilon \simeq \Delta$,
we see that this approximation is justified provided $G^{\rm i}, G^{\rm o} \gg G_{\rm Q} \Delta \nu_0 {\cal V}_k$, or,
equivalently, the Thouless energy of the structure $E_{\rm Th} \simeq (G/G_{\rm Q})/(\nu_0 {\cal V}_k)$
exceeds by far the energy scale $\Delta$.
In terms of the scattering approach outlined below,
it implies that one can neglect the energy dependence of the scattering matrix of
the nanostructure at the energy scale $\simeq \Delta$.

To summarize, the energy spectrum of ABS under our assumption spreads from $0$ to $\Delta$,
and immediately depends on three superconducting phases and on the ratio of inner and
outer conductances $G^{\rm i}/G^{\rm o}$. Below we investigate the details of this dependence.
Naturally, the semiclassical approach cannot give the exact positions of ABS levels:
rather, it gives a smooth energy-dependent density of ABS in their quasicontinuous spectrum.

We derive a convenient and general relation between the number of ABS $N(\epsilon)$ in
the energy interval $[0,\epsilon]$ and the extremal value of the action ${\cal S}$,
\begin{equation}
N(\epsilon) = \frac{1}{2\pi} {\rm Im} {\cal S}(\epsilon).
\label{eq:Neps}
\end{equation}

\subsection{Scattering matrix description}
\label{subsec:scatteringmatrix}

\begin{figure}
\includegraphics[width=0.9\columnwidth]{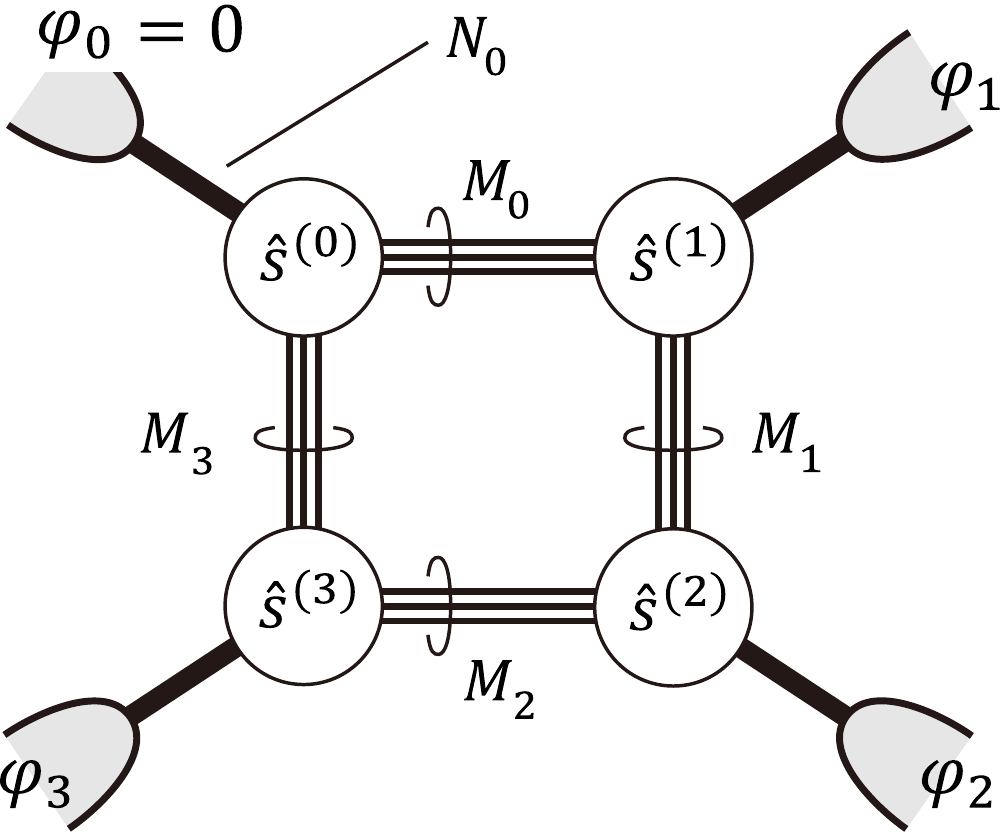}
\caption{
The scattering matrix of the 4T-ring (dimension $\sum_k N_k \times \sum_k N_k$) is
composed of the random unitary matrices $\hat{s}^{(i)}$  of the chaotic cavities.
The dimension of such a matrix is $(N_i +M_i +M_{i-1}) \times (N_i +M_i +M_{i-1})$.
}
\label{fig:model}
\end{figure}

We can evaluate the ABS energies in our 4T-ring from the normal-state
scattering matrix of the device. Here, we rely on Beenakker's determinant equation~\cite{Beenakker}
\begin{equation}
\det \left( e^{i 2\chi} - \hat{S} (\vec{\varphi},E) \right) =0,
\label{eq:Beenakker}
\end{equation}
where the unitary matrix
\begin{equation}
\hat{S} (\vec{\varphi},E) = e^{i \hat{\varphi}} \hat{s}_{\rm h} (E) e^{-i \hat{\varphi}} \hat{s}_{\rm e} (E).
\label{eq:Smatrix}
\end{equation}
incorporates the processes of Andreev reflection in the leads and normal reflection from the device.
$\hat{s}_{\rm e,h} (E)$ are electron and hole scattering matrices in the normal region.
Those are related by  $s_{\rm h} (E) = - \hat{g} \hat{s}_{\rm e}^{\rm *} (-E) \hat{g}$ with $\hat{g} = -i \hat{\sigma}_y$.
$\hat{\sigma}_y$ is a Pauli matrix.
In this Article, we disregard the effects of magnetic field and spin-orbit interaction, thus disregarding
the spin degree of freedom in $\hat{s}_{\rm e,h} (E)$.
$e^{\pm i \hat{\varphi}}$ is a diagonal matrix with $\hat{\varphi} = {\rm diag}(\varphi_0,\varphi_1,\varphi_2,\varphi_3)$
that accounts for Andreev reflection from the corresponding leads.
This form of Beenakker's equation relies on the assumption of the same material for all of the superconducting leads, $\Delta_i = \Delta$.
The Andreev reflection phase $\chi$ is immediately related to energy via $\chi = {\rm arccos} (E/\Delta)$.
Since we consider the limit of a short structure, the scattering matrix $\hat{s}_{\rm e}$ is independent of energy $E$.
The same applies to $\hat{S}$, and the energies of the ABS are readily expressed through the eigenvalues $S_i$ of $\hat{S}$,
$\exp(2 i \chi(E_i) ) = S_i$.

Thus, the normal scattering matrix $\hat{s}_{\rm e}$ determines the Andreev spectrum.
Let us  establish this scattering matrix for the 4T-ring.
It is composed from the scattering matrices of the individual cavities as shown in Fig.\ \ref{fig:model}.
A cavity scattering matrix  $s^{(i)}$ ($i = 0,1,2,3$) describes the scattering between $N_i$ channels
coming from/going to the superconducting terminal $i$,
$M_i$ channels coming from/going to the inner QPC $i$,
and $M_{i-1}$ channels coming from/going to the inner QPC $i-1$.
For example, $\hat{s}^{(0)}$ permits the following block separation corresponding to these channel groups,
\begin{equation}
\left( \begin{array}{c}
b_0 \\
c_{10} \\
c_{30}
\end{array} \right) = \hat{s}^{(0)}
\left( \begin{array}{c}
a_0 \\
c_{01} \\
c_{03}
\end{array} \right)
=\left( \begin{array}{ccc}
r_{00}^{(0)} & t_{01}^{(0)} & t_{03}^{(0)} \\
t_{10}^{(0)} & r_{11}^{(0)} & t_{13}^{(0)} \\
t_{30}^{(0)} & t_{31}^{(0)} & r_{33}^{(0)}
\end{array} \right)
\left( \begin{array}{c}
a_0 \\
c_{01} \\
c_{03}
\end{array} \right),
\end{equation}
where $a_0$ and $b_0$ are the vectors of incoming and outgoing electron amplitudes in the lead $0$, respectively,
while $c_{ij}$ are the vectors of the wave amplitudes going  from the cavity $j$ to the cavity $i$ inside the ring.
Thus, $\hat{s}^{(0)}$ is a $(N_0 + M_0 + M_3) \times (N_0 + M_0 + M_3)$ matrix.
We obtain $\hat{s}_{\rm e}$ by combining $\hat{s}^{(i)}$. To make the combination explicit,
we introduce vectors $\vec{a}$, $\vec{b}$, and $\vec{c}$ as follows:
\begin{equation}
\vec{a} =
\left( \begin{array}{c}
a_0 \\
a_1 \\
a_2 \\
a_3
\end{array} \right), \hspace{2mm}
\vec{b} =
\left( \begin{array}{c}
b_0 \\
b_1 \\
b_2 \\
b_3
\end{array} \right) , \hspace{2mm}
\vec{c} =
\left( \begin{array}{c}
c_{10} \\
c_{21} \\
c_{32} \\
c_{03} \\
c_{01} \\
c_{12} \\
c_{23} \\
c_{30}
\end{array} \right).
\end{equation}
In $\vec{c}$, four upper (lower) components correspond to clockwise (counterclockwise) propagation,
as shown in Fig.\ \ref{fig:model}.
A a complete unitarity matrix of the size $K \times K$, $K \equiv \sum_k (N_k + M_k + M_{k-1})$,
that relates these amplitudes is separated into the following blocks:
\begin{equation}
\left( \begin{array}{c}
\vec{b} \\
\vec{c}
\end{array} \right)
=
\left( \begin{array}{cc}
\hat{X} & \hat{Z} \\
\hat{Y} & \hat{W}
\end{array} \right)
\left( \begin{array}{c}
\vec{a} \\
\vec{c}
\end{array} \right),
\label{eq:fullsmatrix}
\end{equation}
where $\hat{X}$, $\hat{Y}$, $\hat{Z}$, and $\hat{W}$ are given by the elements of $\hat{s}^{(i)}$.
$\hat{X}$ consists of the reflection matrix from and to the channels in the leads while
$\hat{Y}$ ($\hat{Z}$) corresponds to the transmission matrix from the leads (the ring) to
the ring (the leads). The matrix  $\hat{W}$ describes reflection and transmission in the ring.
By eliminating $\vec{c}$ from Eq.\ (\ref{eq:fullsmatrix}), the scattering matrix of the 4T-ring 
defined as $\vec{b} = \hat{s}_{\rm e} \vec{a}$ is reduced to
\begin{equation}
\hat{s}_{\rm e} = \hat{X} + \hat{Z} \frac{1}{1-\hat{W}} \hat{Y}.
\label{eq:4Nsmatrix}
\end{equation}
The size of $\hat{s}_{\rm e}$ is $\sum_{k} N_k\times \sum_{k} N_k$.

The numerical procedure to determine the spectrum of ABS for a given realization of disorder in the 4T-ring could be as follows:
We pick up the $\hat{s}^{(i)}$ for each cavity from the circular ensemble of time-reversible scattering matrices and
form $\hat{s}_{\rm e}$ by making use of Eq.\ (\ref{eq:4Nsmatrix}). For a certain choice of $\varphi_k$,
we form $\hat{S}$ by employing Eq.\ (\ref{eq:Smatrix}) and then diagonalize $\hat{S}$ and
deduce the corresponding ABS energies.

We actually follow all these steps except picking up $\hat{s}^{(i)}$ from the circular ensemble.
We form these matrices in an equivalent but different way that provides numerical efficiency and
has essential physical significance for understanding the properties of the 4T-ring.

We outline this way by concentrating on one of the $\hat{s}^{(i)}$ matrices.
For briefness, we identify $N\equiv N_i$, $2M \equiv M_{i} + M_{i-1}$ and assume $N >2M$.
The matrix $\hat{s}^{(i)}$ is a random $(N+2M) \times (N+2M)$ unitary matrix.
However, $N-2M$ channels on the terminal side of the cavity are completely redundant.
Owing to the mismatch of the number of channels on the terminal and ring sides,
these channels are completely reflected from the cavity not playing any role in the formation of the ABS.
Therefore, we can reduce the matrix dimension by considering only $2M$ channels in the lead.
The resulting $4M \times 4M$ matrices are best presented in terms of
the {\it transmission eigenvalues} from the terminal to the ring side (or back)~\cite{brouwer:96}.

Introducing  a diagonal matrix with $2M$ transmission eigenvalues for the cavity $i$,
$\hat{T}^{(i)} = {\rm diag} (T^{(i)}_1, T^{(i)}_2, \cdots ,T^{(i)}_{2M})$.
we represent $\hat{s}^{(i)}$ as
\begin{equation}
\hat{s}^{(i)} =
\left( \begin{array}{cc}
 \hat{V}^{(i)\prime} &                 \\
                           & \hat{V}^{(i)}
\end{array} \right)
\left( \begin{array}{cc}
-\sqrt{ 1 - \hat{T}^{(i)} } & \sqrt{ \hat{T}^{(i)} }      \\
 \sqrt{ \hat{T}^{(i)} }       & \sqrt{ 1 - \hat{T}^{(i)} }
\end{array} \right)
\left( \begin{array}{cc}
 \hat{U}^{(i)\prime} &                 \\
                           & \hat{U}^{(i)}
\end{array} \right),
\label{eq:Trepresentation}
\end{equation}
where $\hat{U}^{(i)}$, $\hat{U}^{(i)\prime}$, $\hat{V}^{(i)}$ and $\hat{V}^{(i)\prime}$
are $2M \times 2M$ unitary matrices. The size of vectors $a_i$ and $b_i$ is
reduced to $2M$. The four submatrices found in Eq.\ (\ref{eq:Trepresentation}),
$-\hat{V}^{(i)\prime} \sqrt{ 1 - \hat{T}^{(i)} } \hat{U}^{(i)\prime}$,
$  \hat{V}^{(i)}          \sqrt{ \hat{T}^{(i)} }      \hat{U}^{(i)\prime}$,
$  \hat{V}^{(i)\prime} \sqrt{ \hat{T}^{(i)} }      \hat{U}^{(i)}         $ and
$  \hat{V}^{(i)}          \sqrt{ 1 - \hat{T}^{(i)} } \hat{U}^{(i)}         $,
provide the elements of $\hat{X}$, $\hat{Y}$, $\hat{Z}$ and $\hat{W}$, respectively.
In the presence of time-reversal- and spin-rotation symmetries,
$\hat{V}_i = \hat{U}_i^{\rm T}$ and $\hat{V}_i^\prime = \hat{U}_i^{\prime {\rm T}}$.
For a given choice of transmission eigenvalues, these matrices can be taken from the circular ensemble.

In the limit $N,M \gg 1$ the distribution of the transmission eigenvalues is very specific.
It can be derived by elementary methods~\cite{NazarovBlanter} modelling the cavity with
two ballistic contacts of the conductances $G_Q N$, $G_Q 2M$.
The transmission probability reads
\begin{equation}
\rho (T) = \frac{N+2M}{2\pi} \frac{1}{T} \sqrt{ \frac{T-T_{\rm c}}{1-T} }
\label{eq:Tdistribution}
\end{equation}
for $1>T >T_{\rm c}$, $T_{\rm c} \equiv (N-2M)^2 / (N+2M)^2$ and is $0$ otherwise:
there is no chance for a transmission eigenvalue to be smaller than $T_{\rm c}$.
In practice, this means that this chance is exponentially small,
$\propto e^{-2M}$ and can be safely disregarded for our choices of $M$ and $N$.
So we choose a realization of the transmission distribution by generating random numbers that
obey Wigner-Dyson statistics for their spacings and the distribution (\ref{eq:Tdistribution}).

Such a choice ensures numerical efficiency: we work with matrices of the dimension
$2\sum_k M_k \times 2\sum_k M_k$ rather than with the original dimension. To proceed further, we introduce
\begin{equation}
\hat{U} =
\left( \begin{array}{cccc}
\hat{U}^{(0)} & & & \\
 & \hat{U}^{(1)} & & \\
 & & \hat{U}^{(2)} & \\
 & & & \hat{U}^{(3)}
\end{array} \right),
\end{equation}
and the corresponding structures for $\hat{V}$, $\hat{U}^\prime$, $\hat{V}^\prime$, and $\hat{R} = 1-\hat{T}$.
By using these matrices, we express $\hat{X},\hat{Y},\hat{Z}, \hat{W}$ as
\begin{eqnarray}
\hat{X} &=&-\hat{V}^\prime \sqrt{ \hat{R} } \hat{U}^\prime ,\label{eq:X} \\
\hat{Z} &=& \hat{V}^\prime \sqrt{ 1-\hat{R} } \hat{U} \hat{O}_2 ,\label{eq:Z} \\
\hat{W} &=& \hat{O}_1 \hat{V} \sqrt{ \hat{R} } \hat{U} \hat{O}_2 ,\label{eq:W} \\
\hat{Y} &=& \hat{O}_1 \hat{V} \sqrt{ 1-\hat{R} } \hat{U}^\prime ,\label{eq:Y}
\end{eqnarray}
where we have introduced the matrices  $\hat{O}_{1,2}$ that take care of different ordering of $\vec{c}$ for
$\hat{s}_{\rm full}$ in Eq.\ (\ref{eq:fullsmatrix}) and $\hat{s}^{(i)}$. $\hat{O}_{1,2}$ satisfies
\begin{equation}
\hat{O} \equiv \hat{O}_2 \hat{O}_1 =
\left( \begin{array}{cccccccc}
 0 & & & 1 & & & & \\
 & 0 & & & & & 1 & \\
 & & 0 & & & 1 & & \\
 1 & & & 0 & & & & \\
 & & & & 0 & & & 1 \\
 & & 1 & & & 0 & & \\
 & 1 & & & & & 0 & \\
 & & & & 1 & & & 0
\end{array} \right), \hspace{2mm}
\hat{O}^2 =1.
\end{equation}

As Eqs.\ (\ref{eq:X}) - (\ref{eq:Y}) are applied to $\hat{s}_{\rm e}$ in Eq.\ (\ref{eq:4Nsmatrix}),
we see explicitly that $\hat{U}_i^\prime$ and $\hat{V}_i^\prime$ are irrelevant for the ABS energies evaluated from Eq.\ (\ref{eq:Beenakker}).
This is expected due to equivalence of all channels in a superconducting lead with respect to Andreev reflection.
So we set $\hat{U}_i^\prime = \hat{V}_i^\prime = \hat{1}$ without loss of generality. Finally we obtain 
\begin{eqnarray}
\hat{s}_{\rm e} &=& -\sqrt{\hat{R}}
  + \sqrt{1-\hat{R}} \hat{U} \hat{O}_2
  \frac{1}{ 1 - \hat{O}_1 \hat{U}^{\rm T} \sqrt{\hat{R}} \hat{U} \hat{O}_2}
  \hat{O}_1 \hat{U}^{\rm T} \sqrt{1-\hat{R}} \nonumber \\
 &=& -\sqrt{\hat{R}}
  + \sqrt{1-\hat{R}}
  (\hat{K} + \hat{K} \sqrt{\hat{R}} \hat{K} + \cdots )
  \sqrt{1-\hat{R}} \nonumber \\
 &=& -\sqrt{\hat{R}}
  + \sqrt{1-\hat{R}}
  \hat{K} \frac{1}{1-\sqrt{\hat{R}} \hat{K}}
  \sqrt{1-\hat{R}}
\label{eq:8Msmatrix}
\end{eqnarray}
with $\hat{K} \equiv \hat{U} \hat{O} \hat{U}^{\rm T}$.
The second line in Eq.\ (\ref{eq:8Msmatrix}) becomes useful
when we apply a perturbation calculation in small $\hat{R}$ to the determinant equation (\ref{eq:Beenakker}).

A similar reduction of the matrix size is also possible for the opposite case $N<2M$.
Yet it is slightly more difficult to implement it in numerics so we have not done this.

\section{Overview of the spectrum}
\label{sec:overview}
In this Section, we give a general overview of the ABS spectrum in the 4T-ring.
We start with defining the semiclassical topology that provides understanding of the transitions between
gapped and gapless spectra in the device.
The overall properties of the spectrum depend on a dimensionless parameter that
is the ratio of inner and outer conductances, $G^{\rm i}/G^{\rm o} = M/N$.
We describe the properties of the spectrum in the extreme limits of
small (``open limit'') and big (``closed limit'') values of this parameter.
Next, we present numerical illustrations: those obtained by the semiclassical approach as well as
the results of exact diagonalization of the matrix $\hat{S}$ in Eq.\ (\ref{eq:Beenakker}).

The analysis of the results brings us to an important conclusion regarding the topological origin of
the ``smile" gaps in these superconducting nanostructures, that we formulate in subsection E.
The topological protection in this case is provided by the gap in the transmission distribution given by Eq.\ (\ref{eq:Tdistribution}).
We demonstrate in subsection F that an isolated transmission eigenvalue in this gap
results in isolated ABS inside the smile gaps.

Before going to all these details, let us estimate the total number of ABS in the device, $N_{\rm ABS}$.
For a general scattering matrix, the number of Andreev states (with positive energy) derived from
Eq.\ (\ref{eq:Beenakker}) is one half of the matrix dimension.
In our case, this gives $N_{\rm ABS}=\sum_{i} N_i/2$, $N_{\rm ABS}= 2 N$ if all $N_i$ are the same.
However, this estimation does not work if $N>2M$, in this case, as explained in Section \ref{subsec:scatteringmatrix},
$N-2M$ incoming transport channels are reflected back to the same terminal not participating in the formation of ABS.
Therefore, $N_{\rm ABS}= {\rm min}(2 N, 4 M)$.

\subsection{Semiclassical topology}
Before describing the peculiarities of the 4T-ring spectrum,
let us explain the topological properties of the setup that arise at the semi-classical level.
As mentioned, the matrix voltage at $\epsilon = 0$ can be parametrized with real $\theta, \eta$.
It is instructive to associate this matrix with a unit vector on the surface of a sphere, namely,
in its northern hemisphere, $\theta$ being the latitude counted from the equator, $\eta$ being the longitude.
The density of states at zero energy is given by $\nu_0 \sin \theta$.
Therefore, if the superconducting proximity gap is present in the device,
the matrix voltages should all be precisely at the equator, and parametrized by $\eta$ only.
This is plausible since the matrix voltages $\hat{G}^{\rm S}$ in the superconducting terminals are also at the equator,
their longitudes corresponding to their superconducting phases $\varphi_k$.

\begin{figure}
\includegraphics[width=0.8\columnwidth]{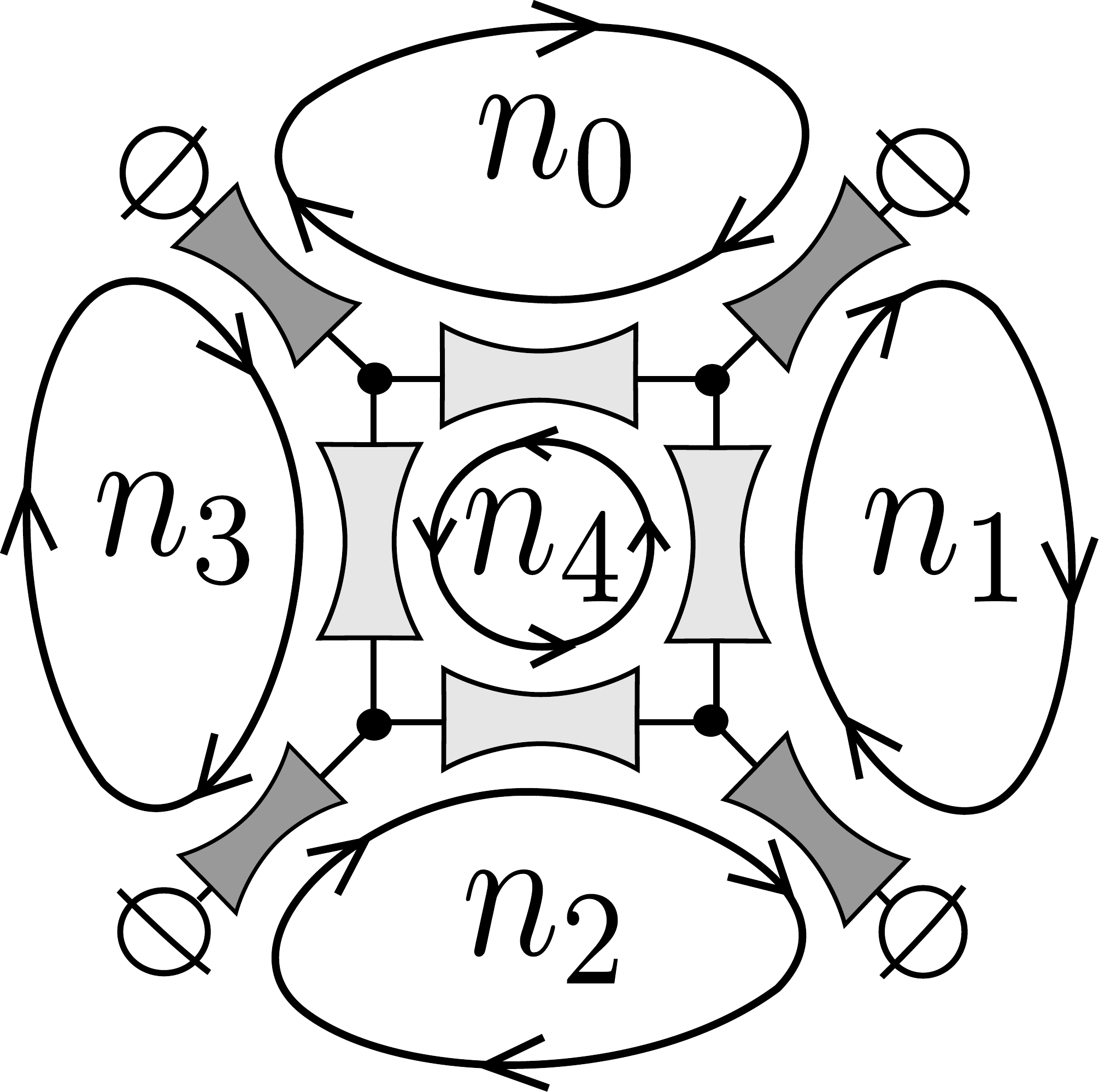}
\caption{
Topological numbers in the 4T-ring.
}
\label{fig:ct-topology}
\end{figure}

Let us show that the possible gapped states of the 4T-ring are distinct in topology and
characterized by 4 independent topological numbers.
In this sense, the gapped states are similar to topologically non-equivalent insulators in
the solid-state physics context~\cite{TopInsulators}.
For 3-terminal structures, the topological analysis of this kind has been suggested and performed in~\cite{PisaNazarov}.

To introduce the topological numbers, let us first concentrate on the central ring of the device.
Similar to the procedure of defining a vortex in Josephson junction arrays~\cite{JArrays},
we sum up the differences of $\eta_i$ over the ring contour projecting each phase difference on $(-\pi,\pi)$ interval.
This defines an integer number $n_4$:
\begin{eqnarray}
2\pi n_4 &=& P(\eta_0-\eta_3)+P(\eta_3-\eta_2)+P(\eta_2-\eta_1)
\\ &+&P(\eta_1-\eta_0);  P(\alpha) \equiv -\pi +2\pi \{\alpha/2\pi +1/2\}, \nonumber
\end{eqnarray}
$\{ \cdots \}$ here denotes the fractional part of a number. The possible values of $n_4$ are $0,\pm 1$.
The configurations of $\eta_k$ with different $n_4$ are topologically distinct since
they cannot be transformed to one another unless one of the phase differences passes $\pm \pi$.
Such a passing, however, would result in a divergent action of the corresponding ballistic connector and
therefore is not realized.

For Josephson arrays, this number indicates the presence of an (anti)vortex in the ring~\cite{JArrays}.
We stress, however, that in our system $\eta_k$ are NOT the phases of superconducting pairing potentials:
there is none in the normal structure under consideration.
Still, the number defined resembles vorticity.

In addition to this, one can define four other topological numbers (Fig.\ \ref{fig:ct-topology})
where a loop is closed through the terminals. In distinction from the previous definition,
the phase difference between the terminals is {\it not} projected on $(-\pi,\pi)$ interval.
For instance,
\begin{equation}
2\pi n_0 = P(\varphi_1-\eta_1) + P(\eta_1-\eta_0) +P(\eta_0-\varphi_0) + \varphi_0 -\varphi_1
\end{equation}  
and $n_{1,2,3}$ are obtained by cyclic permutation of indices.
The justification for such a definition is the fact that nothing special happens to the system
when the difference of the terminal phases passes $\pm \pi$,
so the topological number should experience no change.
A minor disadvantage of the  definition is that topological numbers are not periodic
corresponding to $2 \pi$ periodicity in the 3D space of superconducting phases.
We note that the 5 topological numbers defined are not independent, namely
\begin{equation}
n_4 = \sum_k n_k
\end{equation} 

It is a well-known property of topological insulators that the interface between
two insulators of distinct topology must conduct:
the topology requires such insulators to be separated by a gapless region.
The gapped phases in our device do not have interfaces:
albeit they must be separated by gapless states in parameter space.
We will see this in concrete calculations.

\subsection{Extreme limits: open and closed}

As mentioned, the global properties of the spectrum are determined by
the ratio of the conductances $G^{\rm i}/G^{\rm o} = M/N$. 
First we consider the extreme open limit where the ratio is small,  $M/N \to 0$.
In this limit, a particle coming to a cavity in one of the inner QPC is never reflected back,
but transmits directly to the corresponding superconducting lead.
Upon Andreev reflection in the lead, the particle returns to the same QPC,
transfers it and is Andreev-reflected from another superconducting terminal to return to the same QPC and
complete the cycle. We reckon that all inner QPC in this limit are independent.
The $k$-th QPC hosts a separate bunch of $M_k$ ABS and is biased by the phase difference $\varphi_k - \varphi_{k+1}$.
Therefore, all levels of the bunch have the same energy as in a two-terminal ballistic junction,
\begin{equation}
E= \Delta \cos((\varphi_k-\varphi_{k+1})/2).
\end{equation}
We thus have the case of extreme degeneracy.
In Section IV, we study in detail how this degeneracy is lifted at small but finite values of $G^{\rm i}/G^{\rm o}$.

The result can be derived using the more formal approach of subsection \ref{subsec:scatteringmatrix}.
We observe that in the extreme open limit all transmission eigenvalues in Eq.\ (\ref{eq:Trepresentation})
are concentrated at $T = 1$ ($\hat{R} = 0$) since $T_{\rm c} \to 1$.
Thus, $\hat{s}_{\rm e} = \hat{K}$ and Beenakker's determinant equation (\ref{eq:Beenakker}) becomes
\begin{equation*}
\det \left( e^{i 2\chi} - \hat{S}_0 \right) =0,
\end{equation*}
with
\begin{eqnarray}
\hat{S}_0 &=& \hat{U}^* e^{i \hat{\Phi}} \hat{U}^{\rm T}, \label{eq:S0} \\
\hat{\Phi} &=& {\rm diag}
(\varphi_{01}, \varphi_{03}, \varphi_{12}, \varphi_{10}, \varphi_{23}, \varphi_{21}, \varphi_{30}, \varphi_{32}).
\end{eqnarray}
Here we use $[\hat{U}, e^{i\hat{\varphi}}] = 0$.
Each element of $\hat{\Phi}$ is a phase difference, $\varphi_{ij} \equiv \varphi_i - \varphi_j$,
between adjacent terminals, $j=i\pm 1$.
The eigenvalues of $\hat{S}_0$ are therefore just $\exp(i (\varphi_k - \varphi_{k\pm 1}))$.
Comparing this with $e^{2 i\chi}$ reproduces the above result for the energy.

In the opposite, extreme closed limit, 4 cavities are so strongly coupled as to become
a single cavity characterized by a unitary $4N \times 4N$ matrix.
In circuit-theory description, the system is represented by a single node connected by
ballistic contacts $G^{\rm i}_k$ to the corresponding superconducting reservoirs.
Despite a great simplification, no analytical results for the spectrum can be derived in this limit,
which, as we will see, remains rather complex.
We note, however, that the topological number $n_4$ should be zero in this case,
since $\eta_k$ are the same in all cavities and thus no vorticity can be associated with the ring of the device.

\subsection{Numerics: semiclassics}

\begin{figure}
\includegraphics[width=0.95\columnwidth]{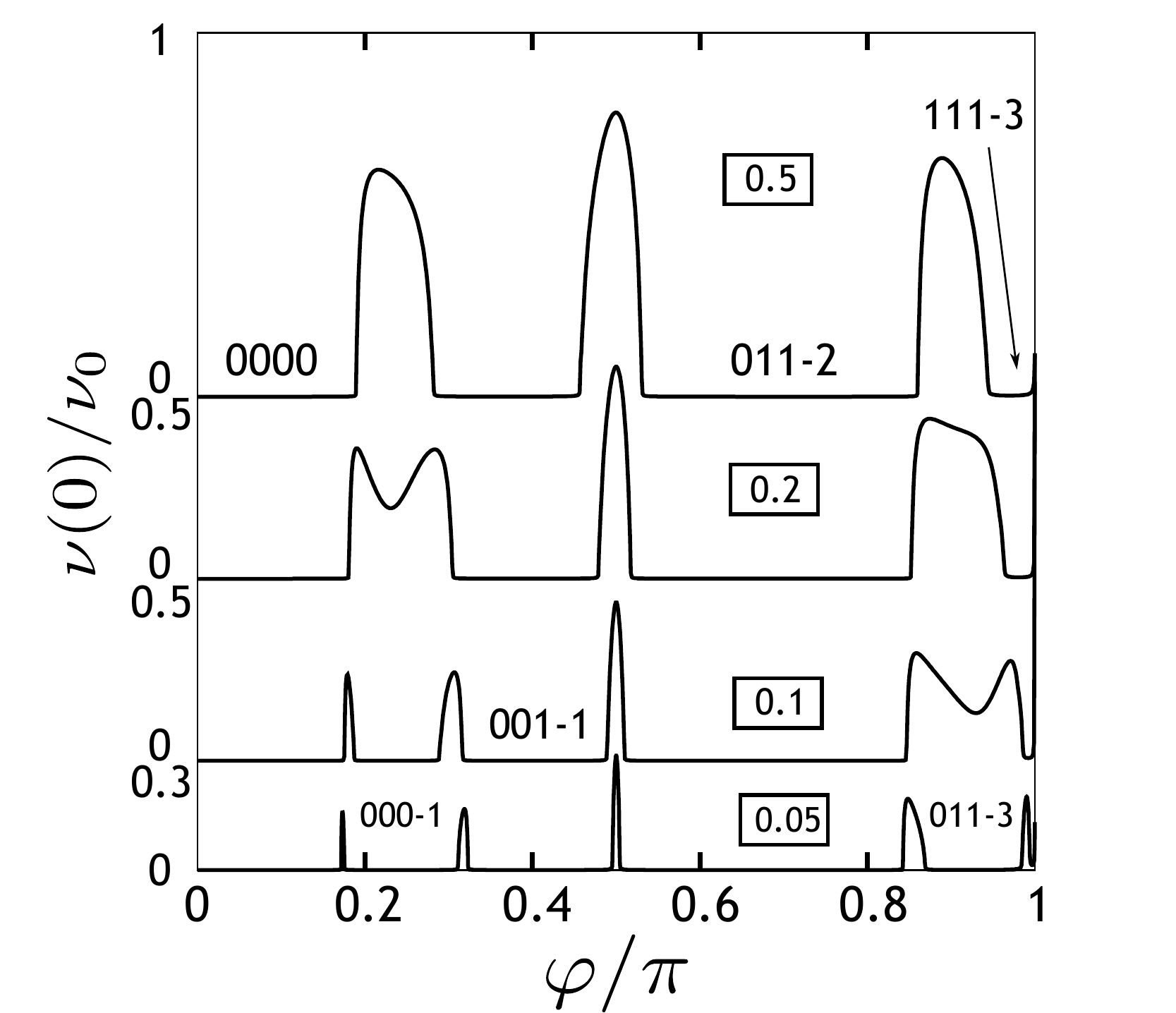}
\caption{
Density of states at $\epsilon=0$ along the line $(\varphi_1, \varphi_2, \varphi_3) = (1,3,6) \varphi$ in the open regime.
The parameter $M/N$ takes values $0.05$, $0.1$, $0.2$, and $0.5$ as indicated by labels in the rectangular frames.
The topological numbers of the gapped states are computed and given in the figure as $n_0 n_1 n_2 n_3$.
}
\label{fig:dosopen}
\end{figure}

\begin{figure}
\includegraphics[width=0.95\columnwidth]{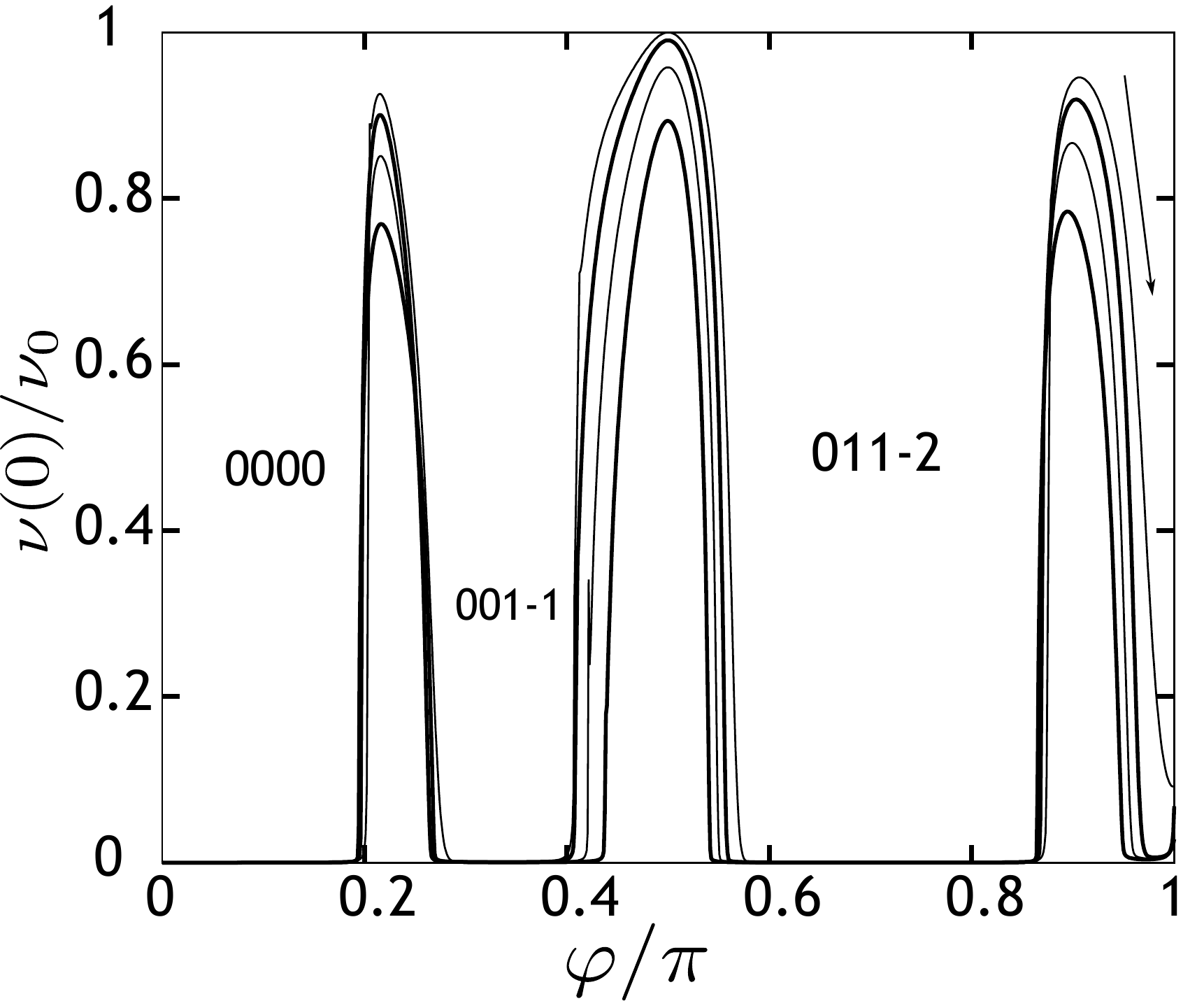}
\caption{
Density of states at $\epsilon=0$ along the line $(\varphi_1, \varphi_2, \varphi_3) = (1,3,6) \varphi$ in the closed regime.
The parameter $M/N$ takes values $1$, $2$, $5$, and $50$ for altering thin and thick curves,
the smaller values of the parameter corresponding to smaller peak d.o.s.
The topological numbers of the gapped states $n_k$ are given.
As expected, no state with $n_4 \equiv \sum_k n_k \ne 0$ occurs in this regime. 
}
\label{fig:dosclosed}
\end{figure}

We present numerical results obtained from the solution of Kirchoff equations corresponding to the action (\ref{eq:action}).
To solve these equations, we employ an iterative algorithm described in Ref.\ \cite{Reulet2003}.

Let us first address the spectral properties at small energy.
Generally, we expect a proximity gap to be induced in the structure.
This would result in a gapped spectrum with no density of states at zero energy.
On the other hand, the analytical results for the open limit show that the ABS come close to zero
any time the phase difference between adjacent terminals approaches $\pi$.

In all illustrations of this article, we explore the spectrum along a line in the 3-dimensional space,
$\varphi_0=0$, $(\varphi_1, \varphi_2, \varphi_3) = (A_1,A_2,A_3) \varphi$.
For most illustrations, we stick to a convenient choice $ (A_1,A_2,A_3)=(1,3,6)$.
In this case, the spectrum is periodic in $\varphi$ with a period $2\pi$ and symmetric with respect to
a transformation $\varphi \to \pi -\varphi$. It suffices to plot the spectrum in the interval $0<\varphi<\pi$.
The phase difference between adjacent terminals approaches $\pi$ for $\varphi = (\pi/6,\pi/3,\pi/2, 5\pi/6)$.

\begin{figure}
\includegraphics[width=0.9\columnwidth]{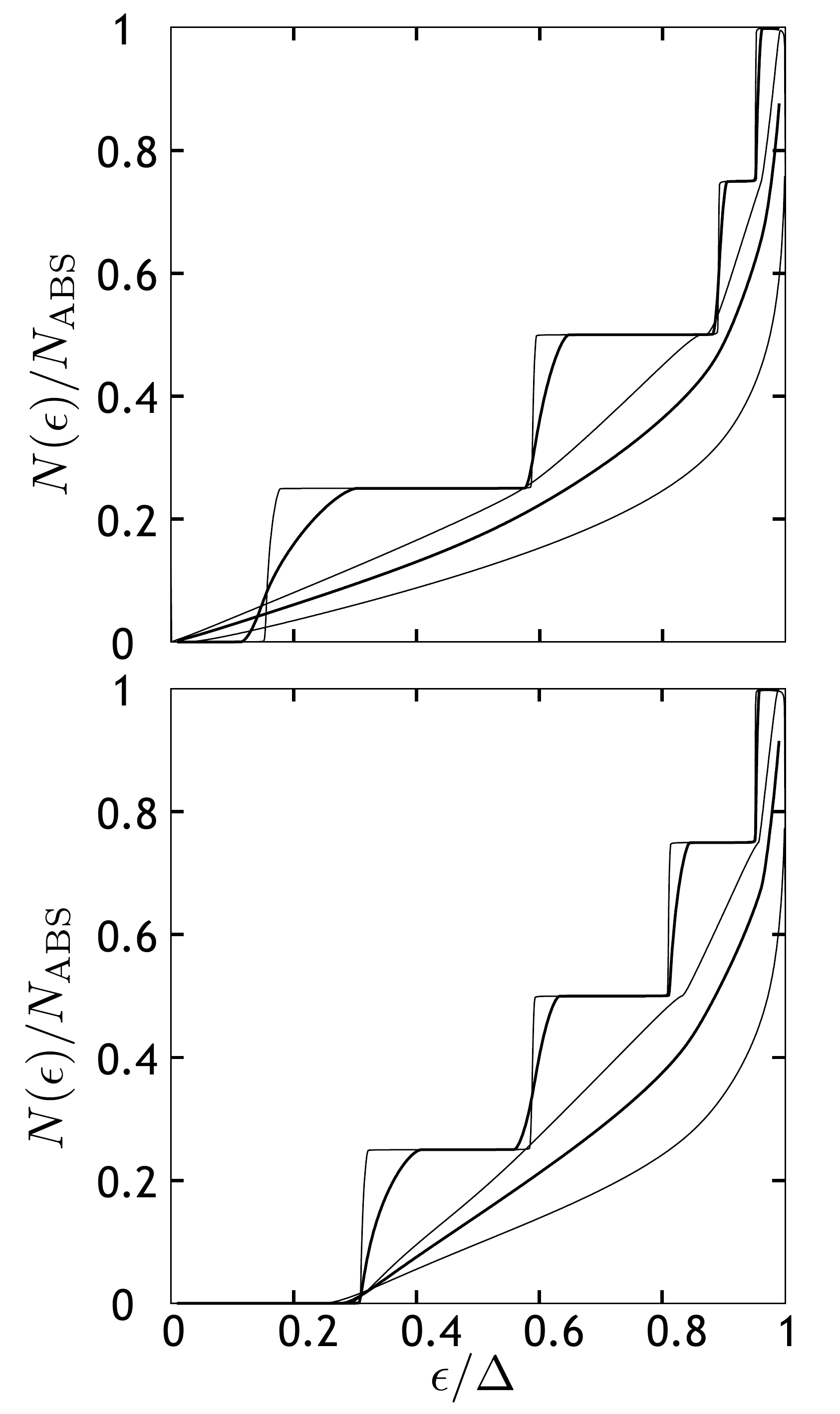}
\caption{
Number of ABS $N (\epsilon)$ versus energy at $\varphi = 0.3 \pi$ (upper panel) and
$0.6 \pi$ (lower panel) on the line $(\varphi_1, \varphi_2, \varphi_3) = (1,3,6) \varphi$ in the open regime.
The parameter $M/N$ takes values $10^{-3}$, $10^{-2}$, $0.1$, $0.2$, and $0.5$ for altering thin and thick curves.
$N_{{\rm ABS}}=4M$.
}
\label{fig:levelsopen}
\end{figure}

In Figs.\ \ref{fig:dosopen} and \ref{fig:dosclosed} we plot the density of states at zero energy versus $\varphi$ for
a representative set of values of $G^{\rm i}/G^{\rm o}$.
The density of states plotted is averaged over the four cavities.

In Fig.\ \ref{fig:dosopen} we concentrate on the open regime, $G^{\rm i}/G^{\rm o} \equiv M/N \le 0.5$.
At small values of the parameter, $\nu(0)=0$ almost everywhere except at narrow peaks around
$\varphi = (\pi/6,\pi/3,\pi/2, 5\pi/6)$ where one of the phase differences between
adjacent terminals approaches $\pi$ pushing the corresponding ABS to zero energy.
We see that these peaks separate gapped states with different topological numbers $n_k$ shown in the figure.
Upon increasing the parameter, the peaks get wider, shift, and sometimes merge so that
some gapped states eventually disappear.
We notice that the disappearing states all have non-zero $n_4 = \sum_k n_k$.
This confirms the expectation that only the states $n_4 =0$ survive in the closed regime.
The density of states slightly increases upon increasing $M/N$.
More interesting details are revealed on the background of these general trends.
For instance, at $\varphi > 0.95$ we see the emergence and stabilization of the gapped state $(111-3)$
that was absent in the limit of vanishing $M/N$.
At $M/N >0.5$ we enter the closed regime.
The peaks get progressively higher and wider yet saturate in both height and width in the extreme closed limit
$M/N \to \infty$ (the curve at $M/N=50$ represents this limit with the accuracy of the plot).
We observe that the state $(111-3)$ disappeares at sufficiently big $M/N$
while most of the gapped states remain in the extreme closed limit.

These figures represent the spectral characteristics at small energy.
Next we consider all the energies of the ABS spanning the interval $0<\epsilon/\Delta<1$.
We compute the total number of ABS $N(\epsilon)$ with energy smaller than $\epsilon$ making use of
Eq.\ (\ref{eq:Neps}) at the same line in phase space taking two values of $\varphi$, $0.3 \pi$ and $0.6 \pi$.
Figure \ref{fig:levelsopen} gives the results in the open regime where $N_{\rm ABS}=4M$.
The curves at small values of $M/N$ are very much step-like, corresponding to the picture of separate,
almost degenerate bunches of levels in each inner QPC.
$N(\epsilon)$ changes within the bunches and has plateaus at $N(\epsilon)=M,2M,3M$
representing the spectral gaps --- ``smile" gaps --- between the bunches. 
We see that upon increasing $M/N$ the curve becomes smoother and the smile gaps eventually disapear,
at least at these particular values of the phases.
For $\varphi=0.3\pi$ this is also associated with the closing of the proximity gap,
while the latter survives at $\varphi=0.6\pi$ up to energies of at least $0.2 \Delta$.

\begin{figure}
\includegraphics[width=\columnwidth]{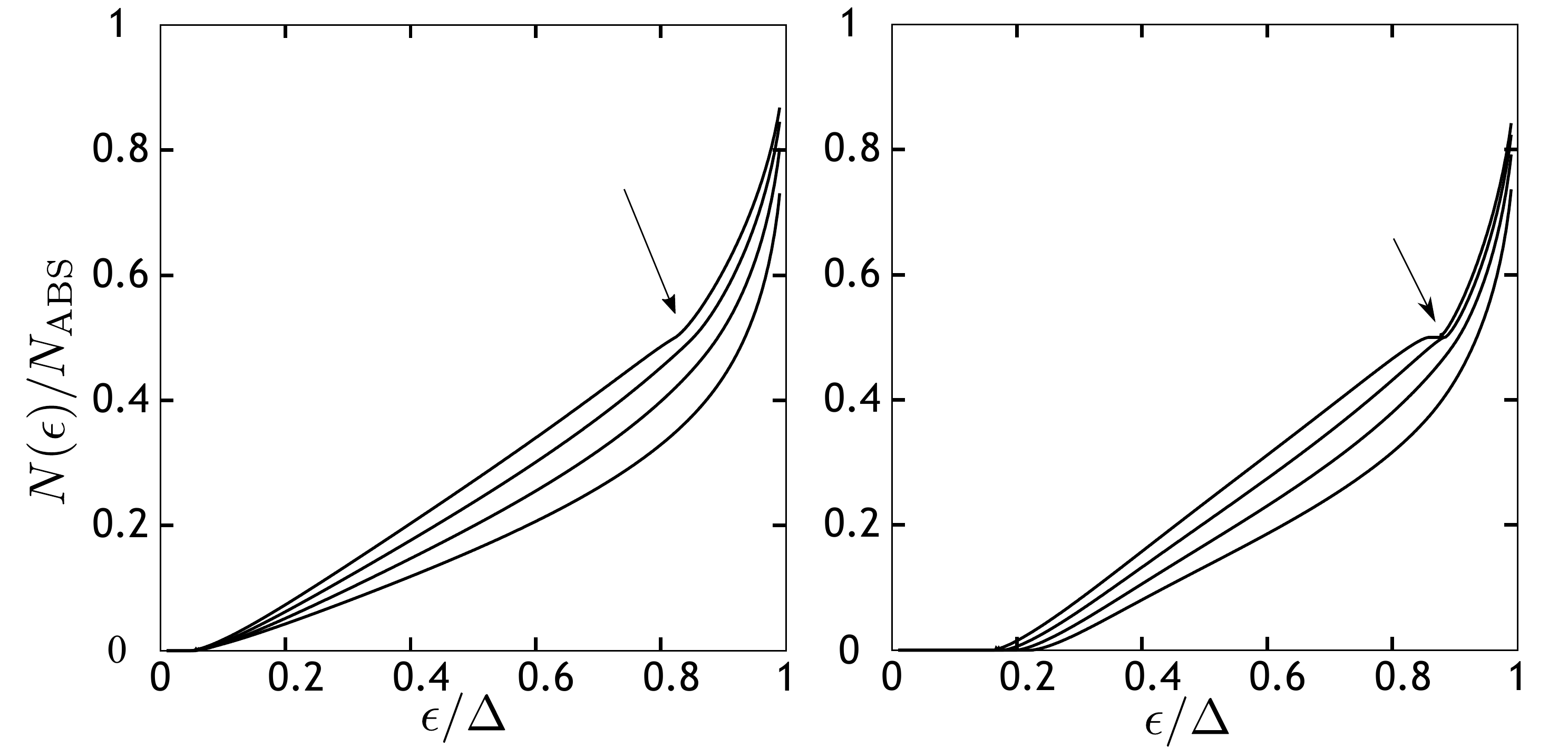}
\caption{
Number of ABS $N (\epsilon)$ versus energy at $\varphi = 0.3 \pi$ (left panel) and
$0.6 \pi$ (right panel) on the line $(\varphi_1, \varphi_2, \varphi_3) = (1,3,6) \varphi$ in the closed regime.
The parameter $M/N$ takes values $1$, $2$, $5$, and $50$ from the lower to the upper curve.
The arrows indicate the formation of the smile gap at $N(\epsilon)= 0.5 N_{\rm ABS}$.
$N_{\rm ABS}=2N$.
}
\label{fig:levesclosed}
\end{figure}

Upon further increase of $M/N$ we enter the closed regime illustrated in Fig.\ \ref{fig:levesclosed}.
It is interesting to note that the smile gaps that have disappeared at moderate $M/N$ reappear at
big values of the parameter, at least at $N(\epsilon)=N=N_{{\rm ABS}}/2$, and the $N(\epsilon)$ curves get sharper.

We explain this with the following consideration.
We note that the 4-terminal system under consideration becomes equivalent to
a 2-terminal one at special symmetry lines in phase space~\cite{YokoyamaNazarov2015}
where the four phases have only two distinct values (upon restricting to an $(-\pi,\pi)$ interval).
Our favorite line $(\varphi_1, \varphi_2, \varphi_3) = (1,3,6) \varphi$ is chosen to cross the symmetry lines.
For instance, at $\varphi = 2\pi/3$, where $\varphi_0, \varphi_2, \varphi_3=0$ while $\varphi_1 = 2\pi/3$.
Thus we deal with $3N$ incoming channels from superconducting terminals at zero phase and
$N$ channels coming from the terminal at non-zero phase.
This restricts the number of ABS to $N$,
to be contrasted with the total number of ABS $N_{{\rm ABS}}=2N$ permitted in the 4-terminal device.
The $N$ non-permitted channels, as we will see in the next subsection, stick to the gap edge.
A smile gap can thus be formed at this special line, and will persist in the vicinity of it.
This is the smile gap seen at $\varphi = 0.6 \pi$ that is close to $2\pi/3$.
A similar effect takes place near $\varphi=\pi/3$ that is not at the symmetry line
but is subject to the same restriction sticking $N$ ABS energies to the gap edge.

\subsection{Numerical results: diagonalization}

\begin{figure}
\includegraphics[width=\columnwidth]{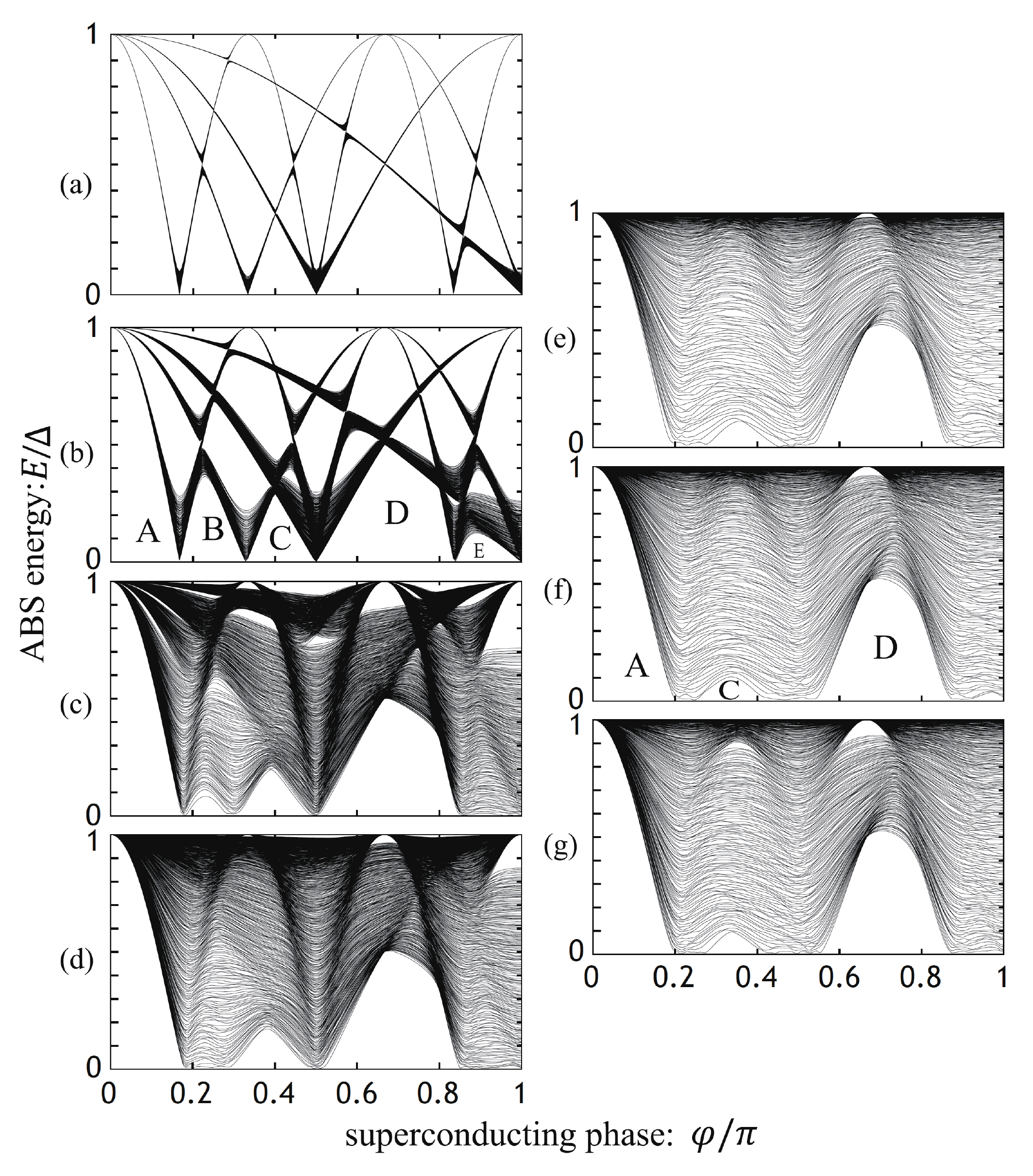}
\caption{
Energy spectrum of ABS in the 4T-ring for various ratios $r = M/N$ of
inner and outer numbers of channels. The superconducting phase is swept on the line
$(\varphi_1, \varphi_2, \varphi_3) = (1,3,6) \varphi$. Only 400 positive ABS energies are shown
in the left panels (a) to (d) and 200 levels in the right panels (e) to (g).
The left panels show the ABS energies in the open regime,
$r = 10^{-3}$(a), $10^{-2}$(b), $10^{-1}$(c), and $0.2$(d).
Here, the number of inner channels in each QPC differ slightly,
$M_0 = 100, M_1 = 120, M_2 = 70, M_3 = 110$. The average is $M = 100$.
Right panels show the ABS energies when the ratio is $r = 1$(e), $2$(f), and $5$(g).
For these cases, $M_i$ are all the same and the number of outer channels is fixed to $N = 100$ in each terminal.
Capital latin letters denote the gapped states with distinct topological numbers,
$A: 0000$,
$B: 000-1$,
$C: 001-1$,
$D: 011-2$,
$E: 011-3$.
}
\label{fig:ABS1}
\end{figure}

To find the ABS energies, we numerically diagonalize the matrix $\hat{S}$ [Eq.\ (\ref{eq:Smatrix})]
for a certain choice of the random scattering matrices in the nodes.
We plot the resulting energies along lines in the 3D space of phases.
For all plots presented, the parameters are chosen to provide $N_{\rm ABS}=400$ bound states in
the energy interval $[ 0,\Delta ]$, except the panels (e), (f), and (g) in the figures where $N_{\rm ABS}=200$.

The number of ABS seems to be sufficiently large for the semiclassical approximation to be valid.
Indeed, we see that the levels mainly follow the behavior of the semiclassical density of states:
there are visible proximity gaps, ``smile" gaps, and a level-bunching in the extreme open limit.
On this background, we also see the signatures of a stochastic parametric dependence typical for random matrix ensembles:
the levels wiggle on the scale of the level spacing $\delta_S \simeq \Delta/N_{{\rm ABS}}$~\cite{Simons},
coming close and further from each other.
The estimations characterizing the stochastic dependence are as follows.
A typical value of the smooth part of the ``velocity" $v=dE/d\varphi$ of a given ABS can be estimated as $\Delta$.
Since the velocity arises from the $N_{\rm ABS}$-component random eigenvector,
the fluctuating part of the velocity can be estimated as
$v_f \simeq v/\sqrt{N_{\rm ABS}} \simeq \Delta/\sqrt{N_{\rm ABS}}$.
From this, a typical scale of the wiggling in the parameter space is estimated as
$\varphi_w \simeq \delta_S/v_f \simeq (N_{\rm ABS})^{-1/2}$.
This is in qualitative agreement with the plots.
Since $v_f \ll v$, the wiggles are most clearly seen around the minima of $v$
where the density of states does not depend much on phase, $v \ll \Delta$.

Next we consider the spectra in more detail. Figure \ref{fig:ABS1} presents the Andreev spectra
along the line $(\varphi_1, \varphi_2, \varphi_3) = (1,3,6) \varphi$ in the 3D space of phases.
Only positive ABS energies are shown.
On the left panel, the spectra are given for small ratios of conductances between inner and outer point contacts,
$M/N <0.5$, where the reduction of the scattering matrix described by Eq.\ (\ref{eq:8Msmatrix}) can be applied.
In Fig.\ \ref{fig:ABS1}(a), the ratio is $M/N = 10^{-3}$.
As we expect from our considerations of the open limit, the levels are grouped into bunches.
In the case of $(\varphi_1, \varphi_2, \varphi_3) = (1,3,6) \varphi$,
all four phase differences between adjacent terminals are different from one oanther,
so we see four bunches in the figure, each encompassing $M_i$ levels. 
The bunches divide the $E-\varphi$-plane into 27 areas.
No isolated level is found inside these areas, implying a ($\varphi$-resolved) gap in the spectrum.
We call the gaps adjacent to zero energy proximity gaps (5 in the Figure) while others are smile gaps.
The width of the bunches increases with increasing ratio, leading to a narrowing and eventually closing of gaps.
At $M/N = 10^{-2}$ [Fig.\ \ref{fig:ABS1}(b)], all 27 gaps in the spectrum are visible,
although the bunch widths are already comparable with the gap size.
In Fig.\ (c) as $M/N = 10^{-1}$, some gaps are evidently closed
while some others are comparable in width with the level spacing.
Most gaps disappear in Fig.\ (d), and, upon crossing to the closed regime, $M/N =1$ [Fig.\ (e)],
the ABS energies are distributed from $E = 0$ to $\Delta$ quasi-continuously at some intervals of $\varphi$.
The density of states at $E = 0$ is finite, and the levels touch the edge of the continuous spectrum at $|E|>\Delta$.
It looks like superconductivity has vanished in these intervals. 
In other intervals, we find the proximity gaps stabilizing for $M/N \geq 1$ [Figs.\ \ref{fig:ABS1}(e)-(g)].
The levels are continuously distributed above the proximity gap.
We also see that upon increasing $M/N$ distinct smile gaps are formed near the symmetry lines,
as explained in Subsection C.

The distinct proximity gaps are associated with the topological numbers given in the figure.
The largest proximity gap corresponds to the $(0000)$ state and occurs at $\varphi =0$
where all the levels stick to the edge of the continuous spectrum.
The second largest gap is about $0.5\Delta$ in either the closed or the open limit corresponding to the state $(011-2)$.

In general, the distribution of levels over energy coincides with the semiclassical predictions of Subsection C.
However, since $N_{{\rm ABS}}$ is still a finite number, there are deviations in the details.
For instance, the semiclassical calculation predicts the proximity gap corresponding to the state $(111-3)$ in a wide interval of $M/N$.
This is not seen in the plots, although the lowest level in the corresponding interval of $\varphi$
deviates from zero more than in other gapped intervals.
The full correspondence is expected to hold at yet larger $N_{{\rm ABS}}$.

\begin{figure}
\includegraphics[width=\columnwidth]{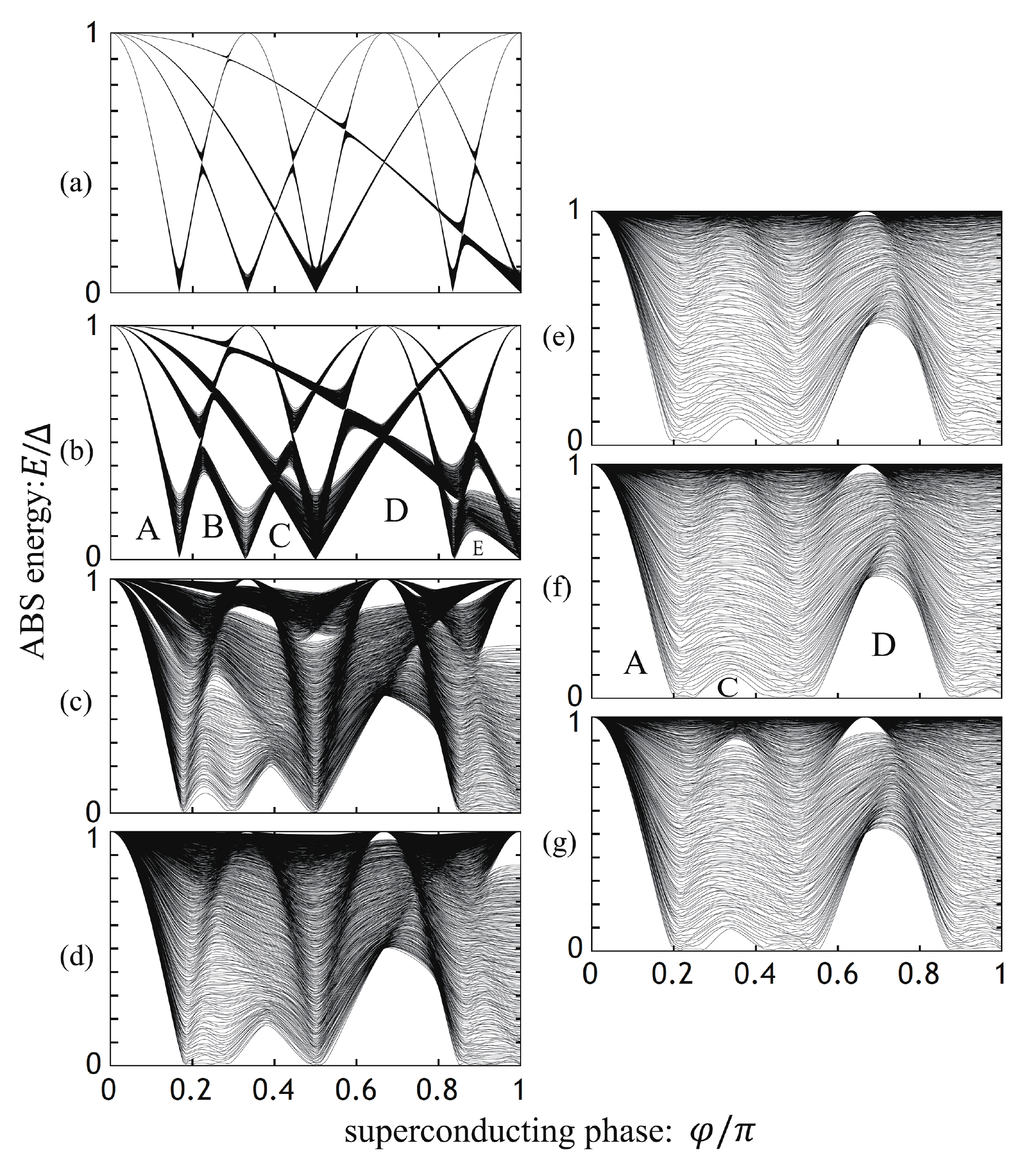}
\caption{
Energy spectrum of ABS in the 4T-ring for various ratios $r = M/N$ of inner and outer numbers of channels.
The superconducting phases satisfy $(\varphi_1, \varphi_2, \varphi_3) = (1,5,10) \varphi$.
The other parameters are the same as in Fig.\ \ref{fig:ABS1}.
Capital Latin letters denote the states with distinct topological numbers,
$A: 0000$,
$B: 000-1$,
$C: 001-1$,
$D: 011-1$,
$E: 011-2$,
$F: 011-3$,
$G: 012-3$,
$H: 012-4$,
$I: 022-4$,
$J: 022-5$.
}
\label{fig:ABS2}
\end{figure}

To estimate the generality of the conclusions, we plot in Fig.\ \ref{fig:ABS2} the spectra
along another line $(\varphi_1, \varphi_2, \varphi_3) = (1,5,10) \varphi$.
The overall picture is significantly more complicated.
In the open limit, four bunches of levels cut the $E-\varphi$ plane  into 50 areas of distinct gaps,
10 of which are proximity gaps characterized by topological numbers.
From these proximity gaps, 5 survive in the closed limit.
The line crosses the symmetry lines at $\varphi=  2\pi /5$ and $\varphi=4\pi /5$
However, the qualitative picture of the spectrum and its evolution with changing $M/N$ is the same.

\begin{figure}
\includegraphics[width=\columnwidth]{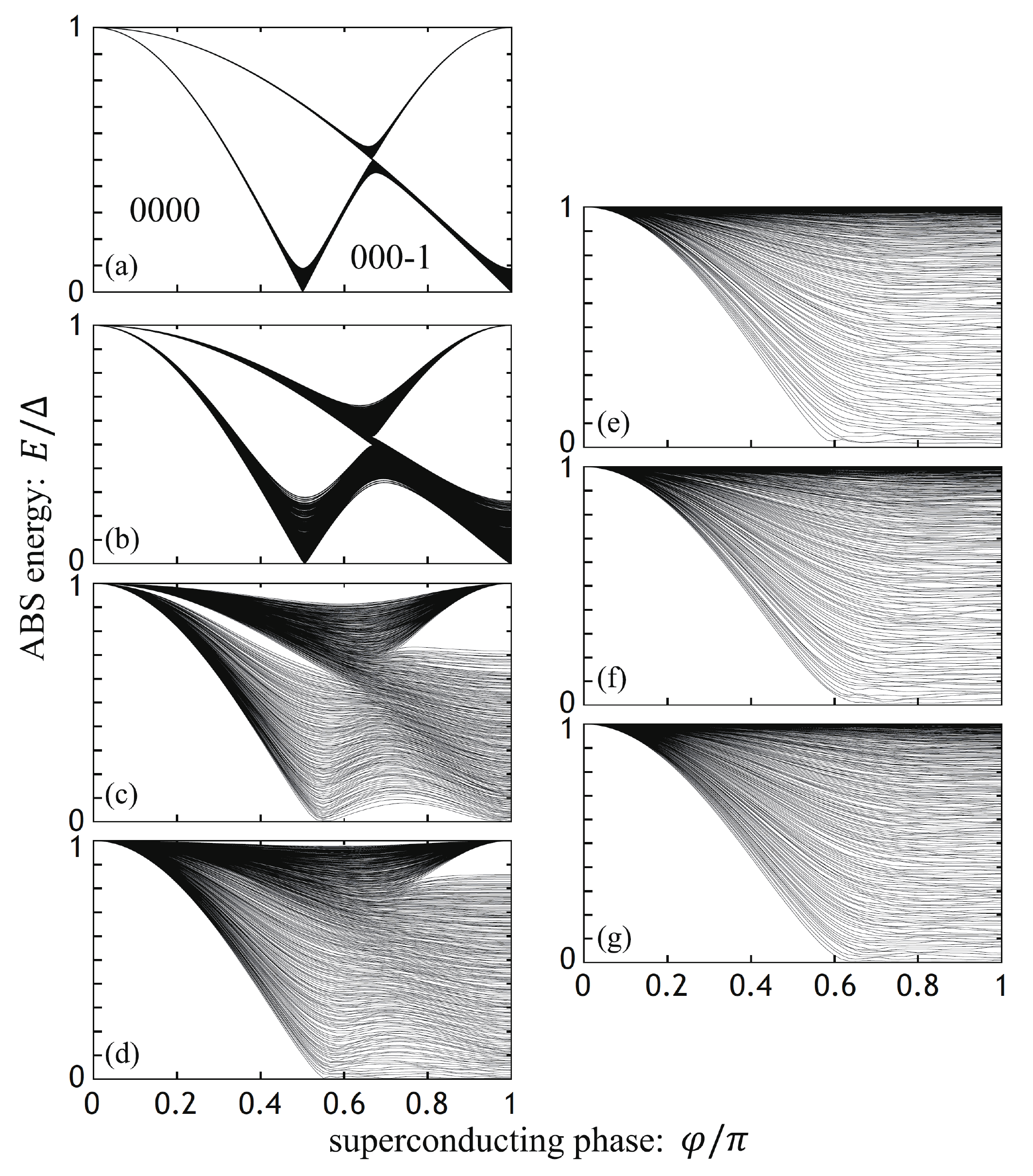}
\caption{
Energy spectra of ABS in the 4T-ring for various ratios $r = M/N$ of inner and outer numbers of channels.
The superconducting phases satify $(\varphi_1, \varphi_2, \varphi_3) = (1,1,2) \varphi$.
The other parameters are the same as in Fig.\ \ref{fig:ABS1}.
The topological numbers of the two gapped states are given in the figure.
}
\label{fig:ABS3}
\end{figure}

A much simpler situation is presented in Fig.\ref{fig:ABS3} for the line $(\varphi_1, \varphi_2, \varphi_3) = (1,1,2) \varphi$.
In this case one of the four bunches is independent of phase $\varphi$ and two are degenerate.
The plane is separated into 5 areas.
In the open limit, there are two proximity gaps with topological numbers $(0000)$ and $(000-1)$.
Since for the second state $n_4 \ne 0$, it does not survive the closed limit disappearing at $M/N \approx 0.17$.

\subsection{Topological protection and origin of the smile gaps}

The smile gaps in disordered systems have been discovered in Ref.\ \cite{SmileGaps}
in the context of two-terminal superconducting structures.
Generally, a smile gap opens in a quasi-continuous spectrum upon changing a parameter:
let us call it $\varphi$, at a splitting point $\varphi_{\rm c}$.
If $\varphi < \varphi_{\rm c}$, all the levels are separated by energies of the order of the level spacing $\delta_S$.
At $\varphi > \varphi_{\rm c}$, two levels separate from each other developing an energy gap
$\Delta E \propto (\varphi-\varphi_{\rm c})^{3/2}$, $\Delta E \gg \delta_S$.
From the point of view of standard theory of spectra in disordered systems,
where close levels are considered to be all alike, the emergence of a smile gap is very confusing.
What actually distinguishes the two levels that separate?
It has been noted in~\cite{reutlinger:14b} that there is a link between
the existence of a gap in the transmission distribution of chaotic cavities and the appearance of the smile gaps.
However, the link only becomes clear in the context of the present device and is explicated here.

We note that the scattering in each of the nodes is described by a scattering matrix of an asymmetric cavity.
A consequence of this asymmetry is that transmission eigenvalues are not distributed in
the whole interval $\lbrack 0, 1 \rbrack$, but there is a minimal transmission eigenvalue $T_{\rm c}$.
$T_{\rm c}$ is a hard boundary for the transmission distribution only in the limit of an infinite number of transport channels.
For a finite number of channels, the random realizations of the scattering matrices permit single transmission eigenvalues below $T_{\rm c}$.
However, these realizations are highly improbable and such transmission eigenvalues appear with exponentially small probability.

The gap in the transmission distribution makes the number of transport channels a {\it relevant} number.
Usually in the context of quantum transport for generic transmission distributions this number is irrelevant~\cite{NazarovBlanter}
since one can always add a channel of vanishing transmission to a connector without
changing the physical properties of the system.
However, this is clearly impossible if the gap is present in the transmission distribution.
This brings us to the conclusion that a 4T-ring setup is characterized by four {\it topological numbers}
that are numbers of the transport channels in the nodes, or, alternatively, in the inner QPC's.
These numbers are topological since they cannot be changed by variations of disorder in the device.

The considerations in the open regime make the link between these topological numbers and the smile gaps obvious.
In the open regime $T_{\rm c}$ is close to $1$ and we find bunches of almost degenerate Andreev levels,
which are separated by large smile gaps.
Since the number of levels in a bunch is $M_i$, the number of levels below a smile gap can be
i. $M_i$;
ii. $M_i+M_j$, $j\ne i$;
iii. $M_i+M_j+M_k$, $j\ne i \ne k, j \ne k$.
This gives 14 distinct possibilities and provides a robust classification of smile gaps.
By virtue of continuity, this classification established in the open limit is valid in the whole space of parameters
where the smile gaps become smaller and eventually close.
In this way, the topological numbers just defined distinguish
the levels that look coequal in a quasi-continuous spectrum.

\begin{figure}
\begin{center}
\includegraphics[width=0.9\columnwidth,angle=0]{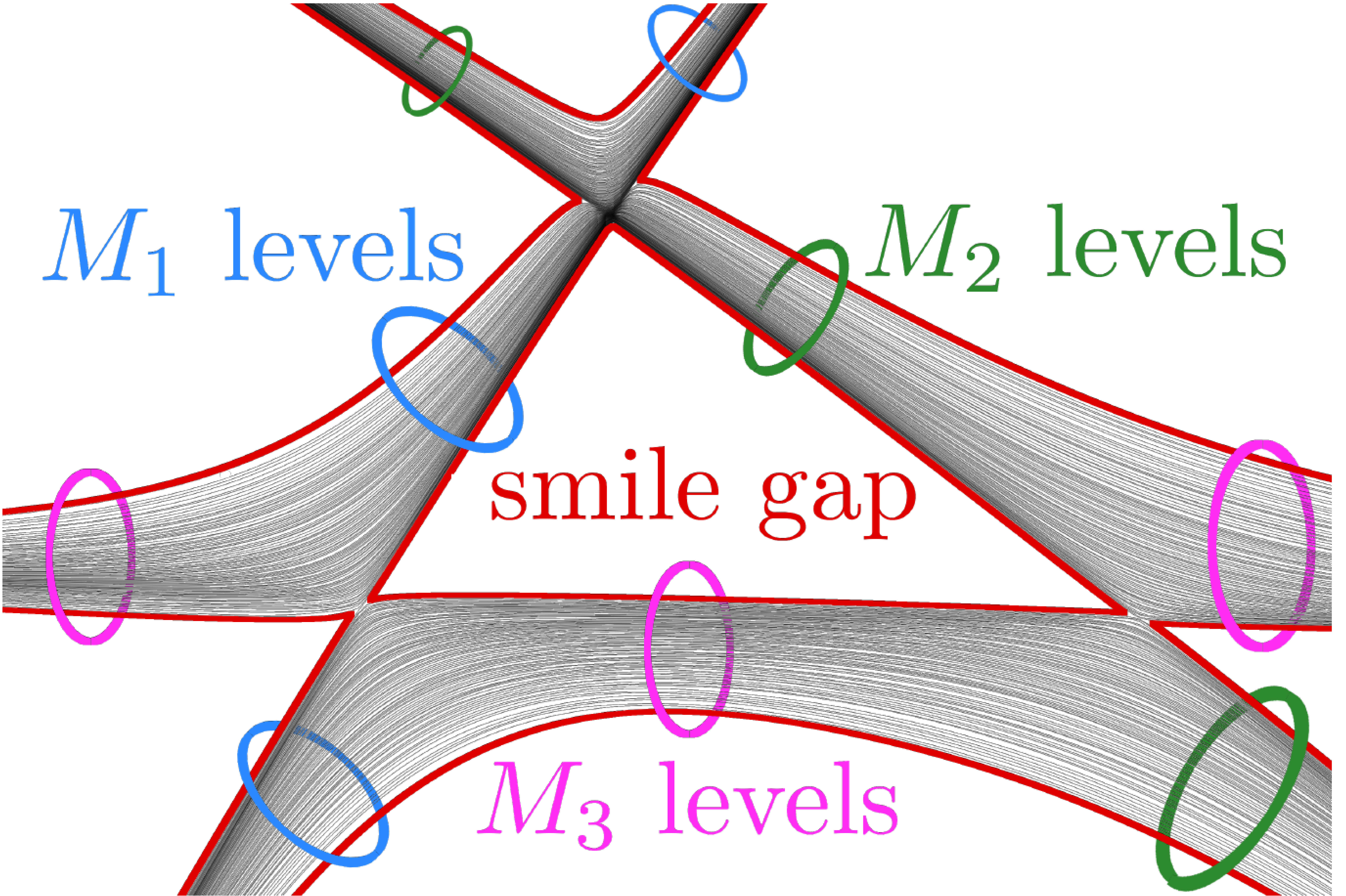}
\end{center}
\caption{(Color online)
Sketch of three bunches that cross at three points surrounding a smile gap.
The bunches have finite width, which is related to the finite width of the transmission distribution in the nodes.
This finite width is given by the thick (red) curves bounding the bunches of Andreev levels.
At each crossing point the number of Andreev levels in each crossing bunch must be conserved.
}
\label{fig:Topo}
\end{figure}

Let us consider in more detail the crossings of bunches to see how the smile gaps are separated from each other.
The bunches have finite width which is related to a small but finite value of $1-T_{\rm c}$.
We find a hard edge on one side with a high level density, where the bunch is confined by
the curve of a level with ideal transmission $T=1$. Since transmission eigenvalues above $T=1$ are not possible,
no random realization of the scattering matrices could break these edges.
On the other side, the levels lie less dense and the boundary of the bunch is defined by
the curve of a level corresponding to $T_{\rm c}$. Thus this edge is no hard but rather soft edge.
The different level densities at the two edges are related to different densities of transmission eigenvalues.
At $T=1$ the transmission distribution diverges, leading to a very dense distribution of Andreev levels,
whereas at $T_{\rm c}$ the distribution remains finite.
The exponential suppression of transmission eigenvalues below $T_{\rm c}$ directly translates into
an exponential suppression of Andreev levels out of the bunches leading to an exponential protection of the smile gaps.
The number of levels in each bunch is constant and equal to the number of
transport modes in the corresponding inner QPC.
These properties are summarized in Fig.\ \ref{fig:Topo},
where the crossings of three bunches, that surround a smile gap, are sketched.
The red lines indicate the finite widths of the bunches.
The number of levels in each bunch must be conserved at each crossing.
If the two bunches have different numbers of levels, some levels have to go straight through
the crossing point in order to assure this.

The gap in the transmission distribution and associated topological protection can be violated
by adding ``by hand" an additional isolated transmission eigenvalue into the gap of the transmission spectrum.
This leads to the violation of the smile gaps: a single Andreev level emerges inside the gap.
We consider this in detail in the next subsection.

\subsection{Stray levels in the smile gaps}


Let us start with the numerical calculation of stray levels. We consider the open limit of small ratio $M/N$,
where Andreev levels come in almost degenerate bunches, which are separated by wide smile gaps.
The minimum transmission $T_{\rm c}$, which is determined by the ration $M/N$, is close to $1$ in this regime.
Andreev levels are mostly localized in one of the inner QPC connecting the neighboring nodes.
We break the gap in the transmission distribution by adding artificially 
only a single transmission eigenvalue. Because of the correspondence of the transmission gap and
the smile gaps this leads to the violation of the smile gaps by a single Andreev level.
While a single Andreev level penetrates into the smile gaps,
all other levels remain in bunches corresponding to a particular ratio $M/N$.
This allows us in principle to study the closing of smile gap by adding levels successively.

\begin{figure}
\begin{center}
\includegraphics[width = \columnwidth,angle=0]{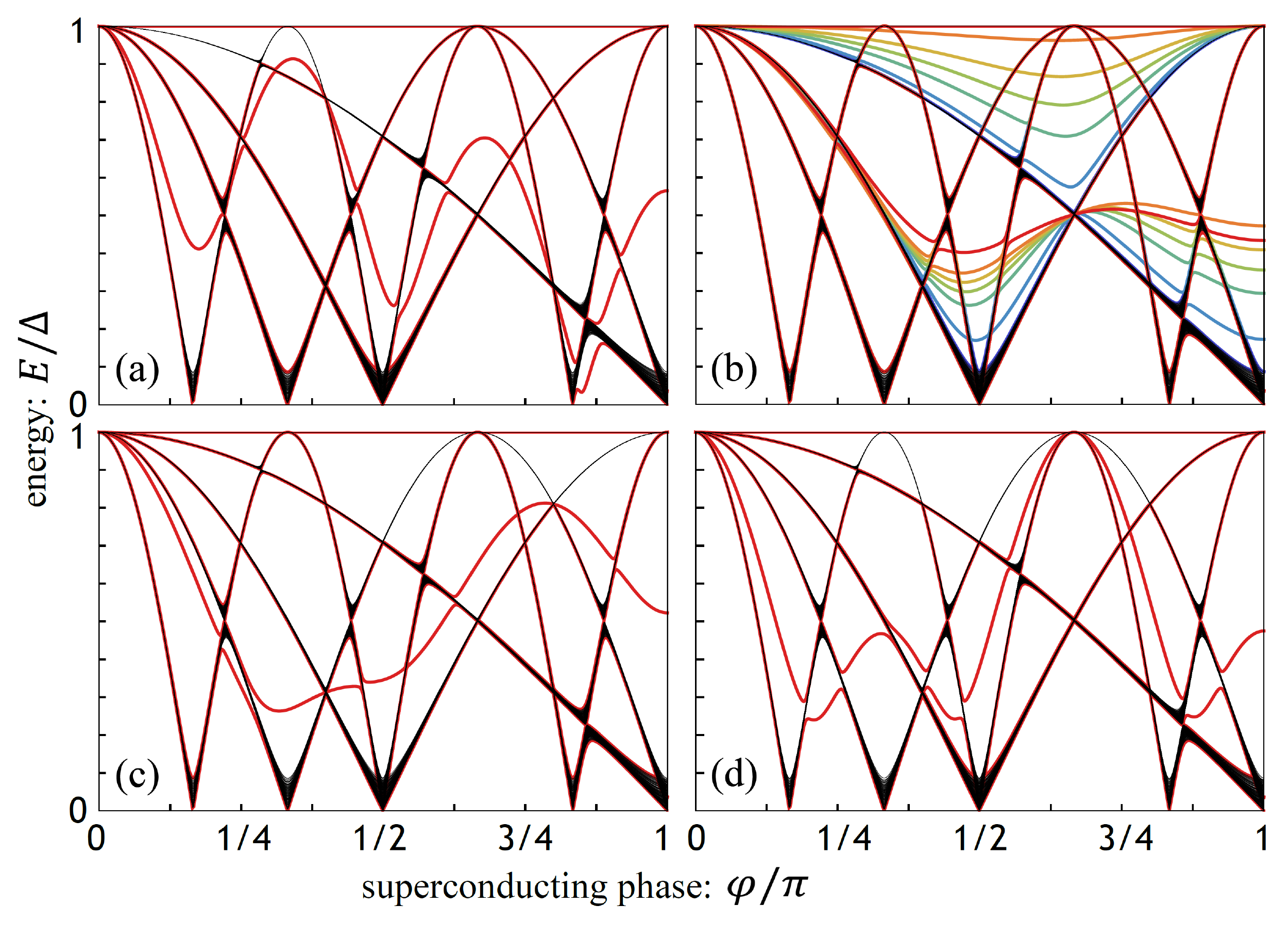}
\end{center}
\caption{(Color online)
Stray levels induced inside the smile gaps by replacing a transmission eigenvalue at
a single node with the value $T_{\rm ext}$ in the transmission distribution gap.
We take equal numbers of internal modes $M_i = M = 100$ and a ratio $M/N=0.001$.
For panels (a), (c), and (d), $T_{\rm ext}=0$ and the eigenvalue is replaced in the nodes 0, 2, and 3, respectively.
In panel (b), stray levels are plotted for a set of $T_{\rm ext}$
changing from $0$ (red curves) to $T_{\rm c}$ (blue curve) in steps of $T_{\rm c}/8$.
The eigenvalue is replaced in node 1.
}
\label{fig:Straylevels1}
\end{figure}

In this calculation, we choose equal numbers of internal modes $M_i = M = 100$ and
the ratio $M/N=1/1000$. In Fig.\ \ref{fig:Straylevels1} a single transmission eigenvalue at
a single node [(a) node 0, (b) node 1, (c) node 2, (d) node 3] is replaced by $T_{\rm ext}$,
while the transmission distributions at the other nodes are not changed.
The superconducting phases are swept along the line $(\varphi_1, \varphi_2, \varphi_3) = (1,3,6) \varphi$.
The panels (a), (c) and (d) show a single stray level where $T_{\rm ext} = 0$ was chosen.
The stray level approximately follows an isolated curve penetrating various gaps and crossing the level bunches.
The curves look like superpositions of simple harmonic functions.
Of course, the single level does not actually cross the bunch:
rather, the level joins the bunch on one side while another level splits from the bunch at the opposite side.
This is clearly seen at all crossings.

In (b), we change $T_{\rm ext}$ from $0$ (the red curve) to $T_{\rm c}$ (the blue curve) in equal steps producing a set of curves.
We see that in fact a single additional transmission eigenvalue produces two isolated ABS.
The reason we see only one level in the panels (a), (c), and (d) is that at $T_{\rm ext} = 0$
one channel is fully reflected, giving rise to a level at $E = \Delta$, which is not visible in the plots.
We see that the positions of the isolated ABS approach the bunches upon $T_{\rm ext} \to T_{\rm c}$.
For $T_{\rm ext} = T_{\rm c}$ the stray level is absorbed by the bunches, and is not visible.

\begin{figure}
\begin{center}
\includegraphics[width = 0.8\columnwidth,angle=0]{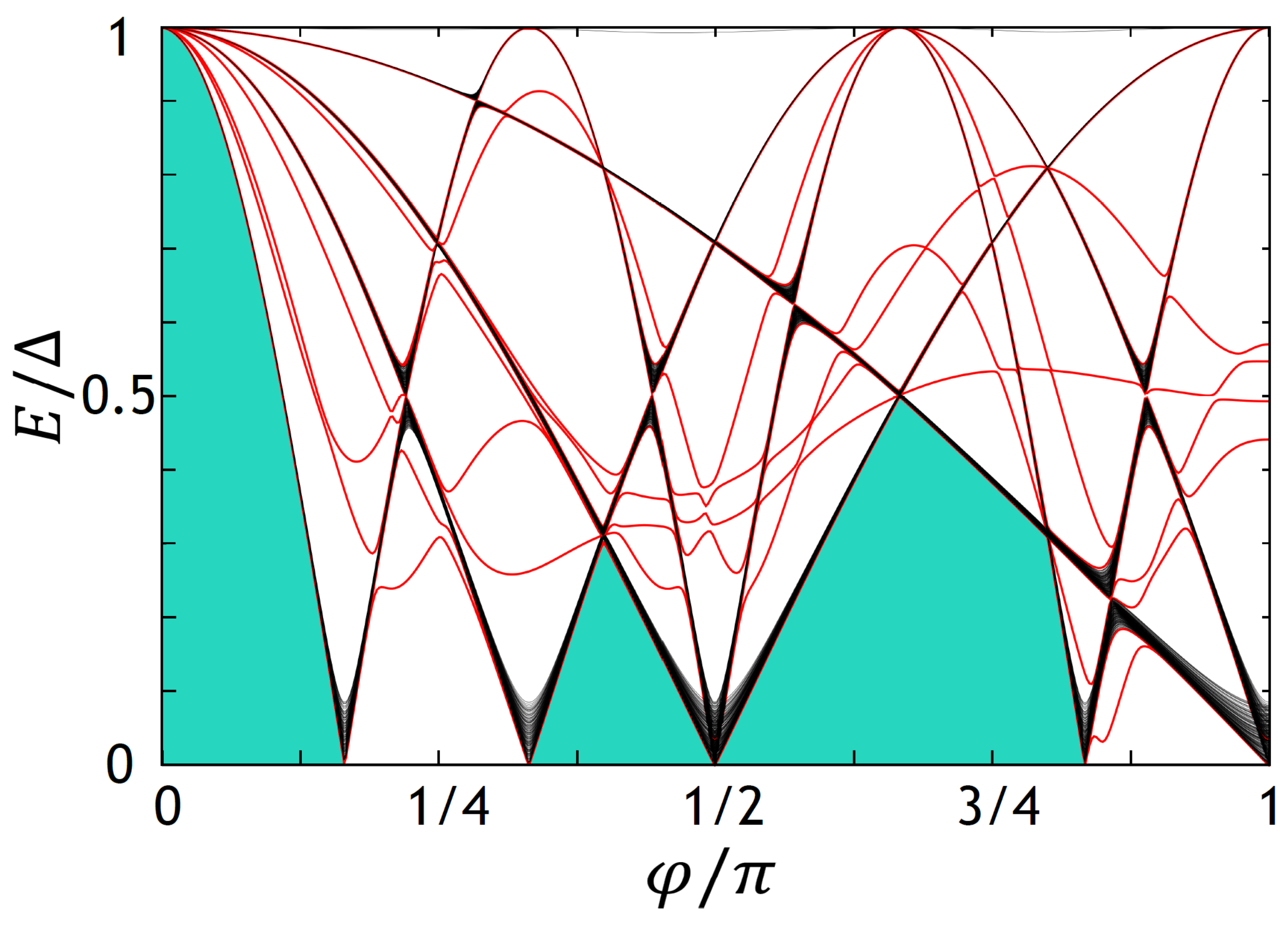}
\end{center}
\caption{(Color online)
Stray levels for extra transmission eigenvalues at all 4 nodes.
At each node, one eigenvalue is replaced to $T_{\rm ext} = 0$.
The number of Andreev levels is the same as that in Fig.\ \ref{fig:Straylevels1} ($4M = 400$).
No stray levels penetrate the proximity gaps marked in green, those survive in the closed limit.
}
\label{fig:Straylevels2}
\end{figure}

Figure \ref{fig:Straylevels2} shows stray levels for the situation where an extra eigenvalue $T_{\rm ext} = 0$ is
replaced at all the nodes. Note that the total number of ABS is $4M = 400$.
We find that none of the stray levels penetrates the three proximity gaps,
which are marked in green in the figure and survive in the closed limit.
This explains their stability upon changing $M/N$ (see Fig.\ \ref{fig:ABS1}).

These numerical results are supported by an analytic calculation of the stray level energies in the extreme open limit $M/N \to 0$.
In this limit all transmission eigenvalues at all nodes are exactly $T = 1$ and
Andreev levels are grouped into degenerate bunches.
We replace a single transmission eigenvalue at node $0$ by $T_{\rm ext} = 0$ and
compute the stray level energy.
In the open limit, the scattering matrix for each node is a $4M \times 4M$ matrix.
Scattering matrices at nodes $i \ne 0$ are given by
\begin{equation}
\hat{s}^{(1,2,3)} =
\begin{pmatrix}
         & \hat{1}_{2M} \\
\hat{1}_{2M} &
\end{pmatrix}.
\end{equation}
Here we put a subscript $2M$ to emphasize the dimension of the identity matrix.
We have chosen the unitary matrices $\hat{U}^{(1,2,3)} = 1_{2M}$, which can be done
without any loss of generality because there is ideal transmission in all channels and phases of
holes cancel those of electrons. For node $0$, the scattering matrix is given by
\begin{equation}
\hat{s}^{(0)} =
\begin{pmatrix}
\hat{1}_{2M} &  \\
          & \hat{U}^{(0)}
\end{pmatrix}
\begin{pmatrix}
-\hat{A}^{(0)} & \hat{B}^{(0)} \\
 \hat{B}^{(0)} & \hat{A}^{(0)}
\end{pmatrix}
\begin{pmatrix}
\hat{1}_{2M} &  \\
         & \hat{U}^{(0)T}
\end{pmatrix}
\label{eq:Node0_S}
\end{equation}
with matrices
\begin{equation}
\hat{A}^{(0)} =
\begin{pmatrix}
\hat{0}_{M} & \; & \; \\
\; & \sqrt{1-T_{\rm ext}} & \;           \\
\; & \;                           & \hat{0}_{M-1}
\end{pmatrix},
\hspace{0.2cm}
\hat{B}^{(0)} =
\begin{pmatrix}
\hat{1}_M & \; & \; \\
\; & \sqrt{T_{\rm ext}} & \;          \\
\; & \;                       & \hat{1}_{M-1}
\end{pmatrix}.
\end{equation}

Without loss of generality we can mix the $T_{\rm ext}$ channel with only a single perfectly transmitting channel,
described by the unitary matrix $\hat{U}^{(0)}$.
This unitary matrix $\hat{U}^{(0)}$ can thus be chosen as a general unitary matrix of the following kind:
\begin{equation}
\hat{U}^{(0)} =
\begin{pmatrix}
\hat{1}_{M-1} & \;                                  & \;                                   & \; \\
\;         & e^{-i\beta} \cos \alpha    & e^{-i\gamma} \sin \alpha & \; \\
\;         & -e^{i\gamma} \sin \alpha & e^{i\beta} \cos \alpha      & \; \\
\;         & \;                                  & \;                                   & \hat{1}_{M-1}
\end{pmatrix}
\end{equation}
where, for a random ensemble of such matrices, $\alpha$, $\beta$ and $\gamma$ are uniformly distributed in the interval $[0, 2\pi]$.
These parameters enter the central block of the matrix $\hat{U}^{(0)}$ that characterizes the channel mixing.
The angle $\alpha$ describes the coupling intensity between outer and inner channels.
To check this, we consider a ($2,1$) block component in $\hat{s}^{(0)}$:
$(\vec{c}_{10}, \vec{c}_{30})^{\rm T} = \hat{U}^{(0)} \hat{B}^{(0)} \vec{a}_0$.
When $\alpha =0$, the matrix $\hat{U}^{(0)}$ is just identical except the phase $\beta$.
If $T_{\rm ext} =0$, one of the channels between node 0 and 1 is disconnected from the terminal 0.
On the other hand, at $\alpha = \pi /2$, one channel between node 0 and 3 is disconnected.

For the present choice of stray levels induced by $T_{\rm ext}$ at node $0$, the node $2$ is irrelevant. 
We can consider a simple scattering problem with a scattering matrix given by Eq.\ (\ref{eq:Node0_S})
that is connected to three superconducting reservoirs 0, 1, and 3.
Beenakker's determinant equation (\ref{eq:Beenakker}) can be reduced in dimension and becomes
$\det(e^{i 2\chi}- \hat{\Phi}^\prime \hat{s}^{(0)*} \hat{\Phi}^{\prime *} \hat{s}^{(0)})$, where
$\hat{\Phi}^\prime = {\rm diag} (e^{i\varphi_0} \hat{1}_{2M}, e^{i\varphi_1} \hat{1}_M, e^{i\varphi_3} \hat{1}_M)$
is a diagonal matrix.
This determinant equation can be solved analytically for a general $T_{\rm ext}$,
however the result is a quite lengthy expression.
We restrict ourselves to the simple case $T_{\rm ext} = 0$.
Two positive-energy solutions are  $E=\Delta$ and
\begin{equation}
\frac{E}{\Delta}=\sqrt{\frac{1+\zeta (\alpha, \varphi_{10}, \varphi_{30})}{2}}
\label{eq:straylevelfunction0}
\end{equation}
with
\begin{eqnarray}
\zeta (\alpha, \varphi_{10}, \varphi_{30}) &=&
\cos^2 \alpha \cos \varphi_{30} + \sin^2 \alpha \cos \varphi_{10} \nonumber \\
&& \hspace{1mm} + \cos^2 \alpha \sin^2 \alpha \left\{ \cos (\varphi_{30} - \varphi_{10}) - 1 \right\}
\label{eq:straylevelfunction}
\end{eqnarray}
Note that $\beta$ and $\gamma$ drop out of the result,
and the stray level energy depends on $\alpha$ only .

\begin{figure}
\begin{center}
\includegraphics[width = 0.8\columnwidth,angle=0]{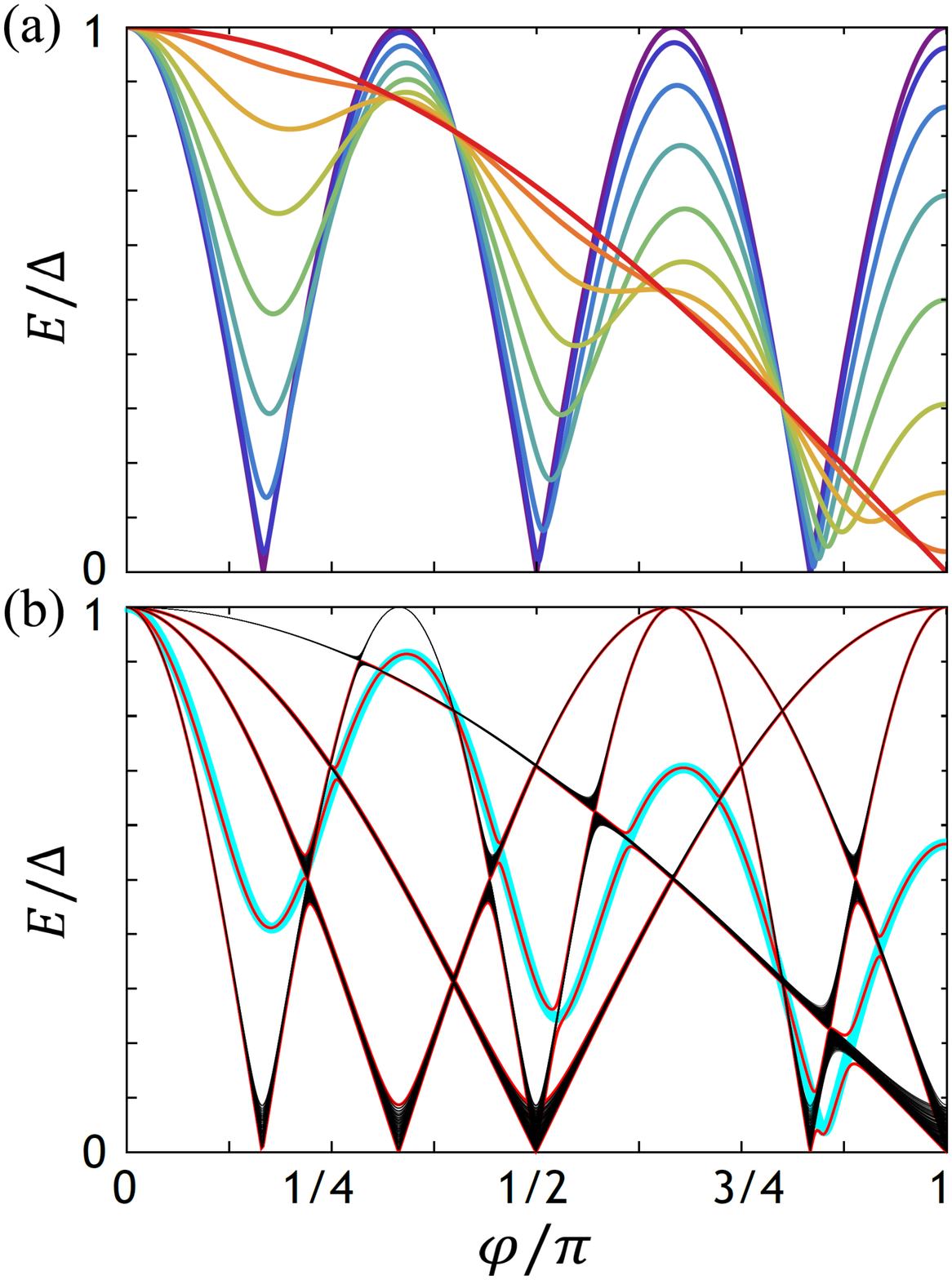}
\end{center}
\caption{(Color online)
Analytical solution given by Eqs.\ (\ref{eq:straylevelfunction0}) and (\ref{eq:straylevelfunction}).
(a) Stray level with $T_{\rm ext} = 0$ at node 0 for a set values of the parameter $\alpha$ ranging from
$0$ to $\pi/2$ in steps of $\pi /16$.
(b) Fit of the analytic expression of the stray level (blue) to the numerically calculated curve (red)
in the case of $\alpha = 0.229\pi$.
}
\label{fig:Straylevels_analytic}
\end{figure}

We plot the analytical solution in Fig.\ \ref{fig:Straylevels_analytic}
along the line $(\varphi_1, \varphi_2, \varphi_3)=(1,3,6)\varphi$.
The figure shows the energy of the stray level for $\alpha$
varying from $0$ (purple) to $\pi/2$ (red) in steps of $\pi/16$.
At $\alpha = 0$, one of the channels in the connector between node 0 and 1 is decoupled from the superconductor 0. 
With this, the stray level energy becomes $|\cos (\varphi_{03} /2)|$ as shown in the figure.
Upon increase of $\alpha$, the stray level energy dependence deviates  from this simple function.
For $\alpha = \pi /2$, only $\varphi_{10}$ is relevant to the level in Eq.\ (\ref{eq:straylevelfunction})
so we  reproduce the $|\cos (\varphi_{10} /2)|$ dependence.
For intermediate values of $\alpha$, the level energy exhibits more complex oscillations.
The behavior of the stray level in Fig.\ \ref{fig:Straylevels1}(a) is reproduced
for $\alpha \approx 0.229\pi$ [Fig.\ \ref{fig:Straylevels_analytic}(b)].

\section{Open limit: crossings and perturbations}
\label{sec:openlimit}

In this section, we investigate in detail the ABS spectrum in the open limit at $M/N \ll 1$,
where the ABS energy levels are grouped in narrow bunches.
Numerical results clearly demonstrate two distinct types of bunch crossings: regular and irregular.
To explain this, and the fine structure of the bunches far from the crossing points,
we develop the perturbation theory up to the second order of $\sqrt{\hat{R}} = \sqrt{1-\hat{T}}$, and apply it.

For numerical illustrations and concrete theory applications, we concentrate on a convenient line 
$(\varphi_1, \varphi_2, \varphi_3) = (1,3,6) \varphi$ in the 3D space of phases.

\subsection{Two types of crossings}

\begin{figure}
\includegraphics[width=0.8\columnwidth]{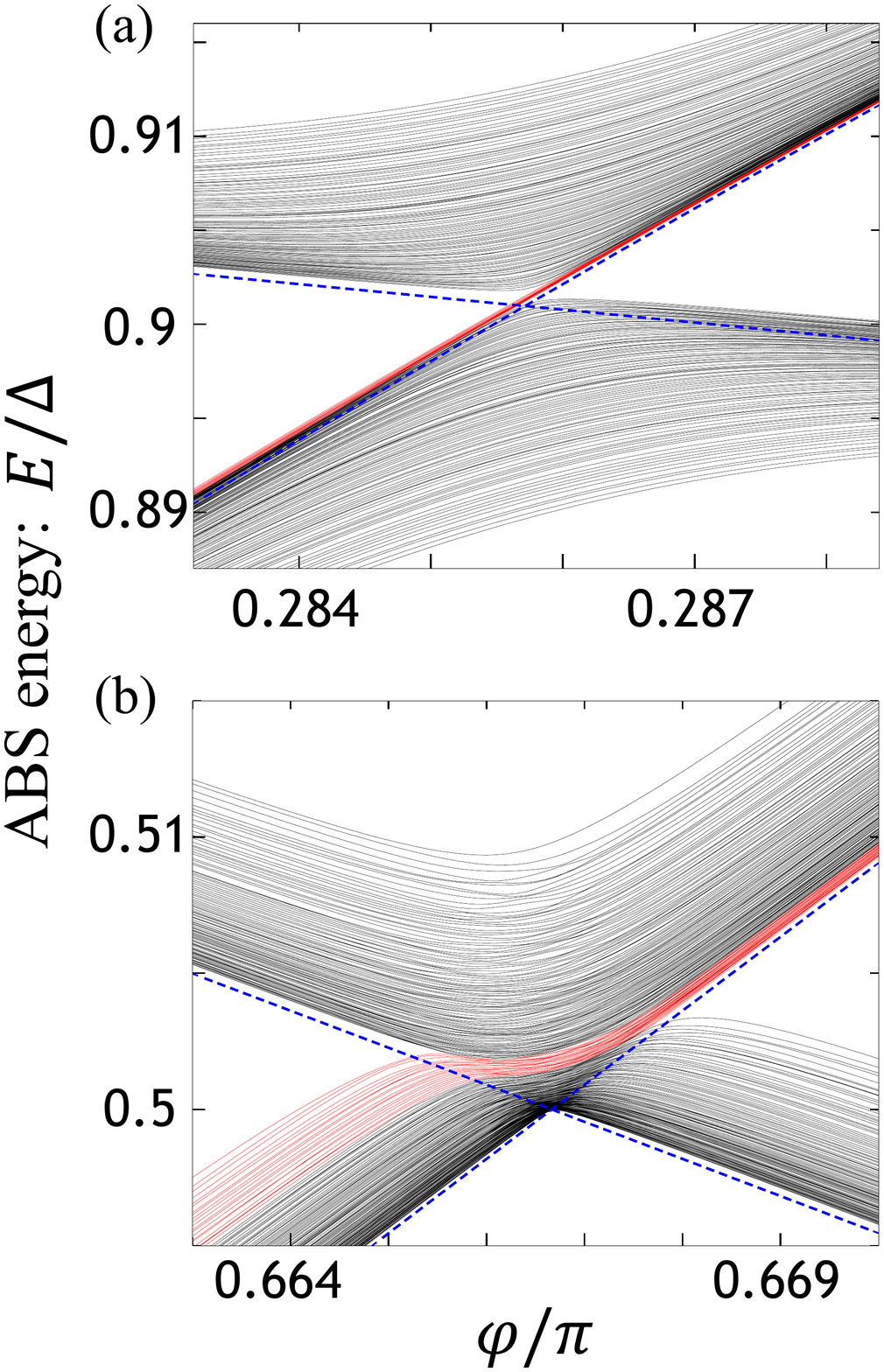}
\caption{(Color online)
Numerics: Two types of bunch crossings in the open limit.
The Andreev levels and assumed parameters are referred from Fig.\ \ref{fig:ABS1}(a).
The parameters for the number of levels in the bunches is $M_0 = 100$, $M_1 = 120$,  and $M_3 = 110$.
The ratio is $M/N = 10^{-3}$. Panels (a) and (b) illustrate a regular  and an irregular crossing, respectively.
Red lines show excess levels, their number being $M_3 - M_0$ in (a) and $M_1 - M_0$ in (b).
The dashed blue lines indicate the reference curves of the bunches, 
$\cos (\varphi_{10}/2)$ and $\cos (\varphi_{03}/2)$ in (a) and $\cos (\varphi_{10}/2)$ and $\cos (\varphi_{12}/2)$ in (b).
}
\label{fig:crossing}
\end{figure}

Figure \ref{fig:crossing} illustates two types of crossings found in our numerical calculations.
In Fig.\ \ref{fig:crossing} (a), we zoom in a crossing of the two bunches
that follow the reference curves $\cos (\varphi_{10}/2)$ and $\cos (\varphi_{03}/2)$
(dashed lines in the plot) in the vicinity of $\varphi_{\rm c} = 2\pi /7$. 
We see that the energy levels are predominantly distributed above and below these curves.
There is a mismatch of numbers of levels in the bunches, $M_3 > M_0$.
$M_3 - M_0$ excess levels exhibit a quasi-linear dependence near the crossing point (red curves),
while $M_0$ pairs of levels exhibit a typical pairwise quasi-hyperbolic level repulsion behavior.
Despite disorder, the phase dependence is very regular in the vicinity of the crossing.
This is an example of a {\it regular} crossing.

Figure \ref{fig:crossing}(b) exemplifies an {\it irregular} crossing.
While the phase dependence of all levels is quasi-linear at some distance from the crossing,
it is obviously irregular in the crossing region conform to expectations for a disordered system.
This is true for the excess levels (red curves) as well. 
No levels are found below the reference lines.
The distribution of the levels exhibits a sharp edge at the lowest lines.

In both cases, the smile gaps are formed on the left and on the right of a crossing.
Figures \ref{fig:crossing} (a) and (b) show a qualitative difference in the gap opening in the vicinity of a crossing $\varphi_{\rm c}$.
For a regular crossing, the gaps open  in a quasi-linear fashion $\Delta_E \propto (\varphi - \varphi_{\rm c})$,
which follows the lines in the extreme open limit.
For the irregular one, the gap closing and opening near the crossing point
follows a law that signifies disorder, $\delta_{\rm gap} \propto (\varphi - \varphi_{\rm 0})^{(3/2)}$,
and the point of the opening $\varphi_{\rm 0}$ is noticeably shifted with respect to $\varphi_{\rm c}$.

There is a clear difference between the two types of crossings.
The perturbation analysis presented below shows that the degeneracy lifting of the regular crossings is dominated by first-order perturbations.
The first order terms vanish for an irregular crossing, so the degeneracy lifting is governed by second-order terms.

\subsection{Perturbation theory}

In this Subsection, we develop a perturbation theory suitable for the open limit $M/N \ll 1$.
In this case, the transmission eigenvalues for
$\hat{s}^{(i)}$ in Eq.\ (\ref{eq:Trepresentation}) are distributed near $1$ ($\hat{R} \sim 0$).
We can thus use the reflection amplitudes as the parameters of the perturbation expansion.
The determinant equation (\ref{eq:Beenakker}) can be rewritten as an eigenvalue equation for a Hermitian matrix:
\begin{equation}
\hat{H}_{\rm eff} (\vec{\varphi}) | \psi \rangle = \tan \chi | \psi \rangle,
\label{eq:detdigen}
\end{equation}
with an effective Hamiltonian given by
\begin{equation}
\hat{H}_{\rm eff} (\vec{\varphi}) \equiv i \frac{1-\hat{S} (\vec{\varphi})}{1+\hat{S} (\vec{\varphi})}.
\label{eq:Hamiltonian}
\end{equation}
To simplify the notation, we drop the ``hat'' symbol for the matrices $K$, $U$, $H$, $S$, and $R$.
We expand  the scattering matrix in $\sqrt{R}$ up to second order,
$S \approx S_0 + S_1 + S_2$, and
substitute the result to the Hamiltonian (\ref{eq:Hamiltonian}):
\begin{eqnarray}
H_{\rm eff} (\vec{\varphi}) &\approx & H_0 + H_1 + H_2,
\label{eq:eff:Hamiltonian} \\
H_0 &=& i \frac{1-S_0}{1+S_0},
\label{eq:H0} \\
H_1 &=& -i2 \frac{1}{1+S_0} S_1 \frac{1}{1+S_0},
\label{eq:H1} \\
H_2 &=& i2 \frac{1}{1+S_0} \left\{ S_1 \frac{1}{1+S_0} S_1 - S_2 \right\} \frac{1}{1+S_0}.
\label{eq:H2}
\end{eqnarray}
$H_1$ and $H_2$ are of the order $\sqrt{R}$ and $R$, respectively.

Before specifying to our setup, we review a general perturbation theory approach for
degenerate levels with perturbation terms of first and second order.
Let $|n_i \rangle$ be an eigenstate of the unperturbed Hamiltonian $H_0 |n_i \rangle = H^{(0)}_n |n_i \rangle$
where all states with the same $n$ are degenerate. 
To first order, the splitting of these energy levels is obtained from diagonalization of a matrix
\begin{equation}
H_{{\rm eff}} =\langle n_i | H_1 | n_j \rangle.
\label{eq:correction1}
\end{equation}

To second order, this degeneracy-lifting matrix is contributed by products of the matrix elements of  $H_1$ and
the elements of  $H_2$:
\begin{equation}
H_{{\rm eff}} =\sum_{m\ne n,k} \frac{ \langle n_i| H_1 |m_k\rangle\langle m_k|H_1 | n_j \rangle}{ H^{(0)}_n - H^{(0)}_m} 
+ \langle n_i | H_2 | n_j \rangle.
\label{eq:correction2}
\end{equation}

For the 4-T ring we expand Eq. (\ref{eq:8Msmatrix}) in $\sqrt{R}$ and arrive at
\begin{eqnarray}
s_{\rm e} \approx K &+& \left( -\sqrt{R} + K \sqrt{R} K \right) \nonumber \\
&& + \left\{ K \sqrt{R} K \sqrt{R} K - \frac{1}{2} (K R + R K) \right\}. 
\end{eqnarray}
Since $S_0$ is given in Eq.\ (\ref{eq:S0}), the unperturbed Hamiltonian is rewritten
as $H_0 = U^* \tan (\hat{\Phi} /2 + m\pi) U^{\rm T}$ with integer $m$.
Its eigenvector is given by $U^* |i \rangle$ with a normalized $|i \rangle$.

In this respect, it is instructive to use an equivalent matrix $G = U^{\rm T} \sqrt{R} U$ as
a perturbation parameter. It satisfies $G^* = G^\dagger$. With this, we obtain
\begin{eqnarray}
U^{\rm T} S_1 U^* &=& e^{i \hat{\varphi}} \hat{O} e^{-i \hat{\varphi}} (\hat{O} G \hat{O} - G^\dagger )
+ e^{i \hat{\varphi}} (\hat{O} G^\dagger \hat{O} - G) e^{-i \hat{\varphi}} \hat{O},
\label{eq:S1} \\
U^{\rm T} S_2 U^* &=& e^{i \hat{\varphi}} \hat{O} e^{-i \hat{\varphi}}
\left\{ \hat{O} G \hat{O} G \hat{O} - \frac{1}{2} (\hat{O} GG^\dagger + G^\dagger G \hat{O}) \right\},
\nonumber \\
&& + e^{i \hat{\varphi}}
\left\{ \hat{O} G^\dagger \hat{O} G^\dagger \hat{O}
- \frac{1}{2} (\hat{O} G^\dagger G + GG^\dagger \hat{O}) \right\} e^{-i \hat{\varphi}} \hat{O},
\nonumber \\
&& + e^{i \hat{\varphi}} (\hat{O} G^\dagger \hat{O} - G)
e^{-i \hat{\varphi}} (\hat{O} G \hat{O} - G^\dagger )
\label{eq:S2}
\end{eqnarray}
Eqs.\ (\ref{eq:S1}) and (\ref{eq:S2})
are used to compute the perturbation corrections.

\subsection{Regular crossings}

In this Subsection, we concentrate on the perturbative corrections that arise from the first order terms in
$\sqrt{R}$ in the effective Hamiltonian (\ref{eq:eff:Hamiltonian}).
In the extreme open limit ($M/N \to 0$), $R$ vanishes resulting in a $M_i$-fold degeneracy
of the levels in the bunch associated with the $i$-th QPC.
Generally, one expects this degeneracy to be lifted already in the first non-vanishing order of the perturbation theory.
This, however, is not the case in our 4T-ring setup.
As a matter of fact, the matrix elements of the first order perturbations vanish,
$\langle n | H_1 | n' \rangle = 0$, for all states $n$,$n'$ that belong to the same bunch.
However, this does not imply that the first-order terms are completely irrelevant:
they play a role in the vicinity of the crossing points of two bunches $j$,$i$ removing
the $M_i+M_j$-fold degeneracy near this point. Here, $H_1$ mixes the levels of different bunches.
In the following, we concentrate on the vicinity of a specific crossing.
The results can be straightforwardly extended to all other crossings of the same type.

We consider the crossing of the bunches following $\cos (\varphi_{10}/2)$ and $\cos (\varphi_{03}/2)$
along the line $(\varphi_1, \varphi_2, \varphi_3) = (1,3,6)\varphi$.
The ABSs corresponding to $\varphi_{10}$ and $\varphi_{03}$ are localized at QPC $0$ and $3$
(and connecting terminals), respectively.
The crossing occurs at $\varphi = 2\pi/7$ as shown in Fig.\ \ref{fig:ABS1} (a).
The effective Hamiltonian including 0th and 1st order terms reads
\begin{displaymath}
H_0 + H_1 = U^* \left \lbrack \tan{(\hat \varphi/2})
- 2i \frac{1}{1+e^{i \hat \varphi}} (U^T S_1 U^*) \frac{1}{1+e^{i \hat \varphi}}\right \rbrack U^T,
\end{displaymath}
where $U^T S_1 U^*$ is defined by Eq.\ (\ref{eq:S1}).
To perform a projection on the subspace of degenerate levels,
it is instructive to subdivide the matrix $G$ in $M_i \times M_j$ blocks
$G^{(k)}_{ij}$, $i$, $j$ being the QPCs adjacent to the node $k$.
The projected Hamiltonian reads 
\begin{eqnarray}
\label{eq:Reg_H}
H_{{\rm eff}} &=&
\begin{pmatrix}
\tan({\varphi_{03} /2}) &  h^* G^{(0)}_{30} \\
h G^{(0)\dagger}_{30}  & \tan({\varphi_{10} /2})
\end{pmatrix}.
\end{eqnarray}
where the diagonal terms are of 0th order and given by the degenerate expressions
$\tan (\varphi_{03}/2)$ and $\tan (\varphi_{10}/2)$
(Note that at $\varphi = 2\pi/7$, $\tan({\varphi_{03} /2}) = \tan({\varphi_{10} /2})$),
while the non-diagonal terms are of the first order and lift the degeneracy. 
Here, $h \equiv h(\varphi_{10},\varphi_{03})$ is given by 
\begin{equation}
\label{eq:H}
h (\varphi_{10},\varphi_{03}) \equiv \frac{\tan (\varphi_{03}/2)}{\cos (\varphi_{10}/2) } e^{i\varphi_{10}/2}.
\end{equation}
$G^{(0)}_{30}$ is a matrix of transmission amplitudes describing the scattering of electrons that 
move from node 3 to node 1 reflecting in the node 0. Note that $G_{30}^{(0)} = t_{31}^{(0)}$.
The eigenvalues of $H_{{\rm eff}}$ are readily expressed in terms of the eigenvalues $g_i$ of
the positively defined matrix $G^{(0)\dagger}_{30} G^{(0)}_{30}$.
Assuming $M_3 > M_0$, we notice $M_3 - M_0$ zero eigenvalues of $g_i$.
In this approximation, this results in $M_3 - M_0$ degenerate levels following the curve $\tan \chi =\tan({\varphi_{03} /2})$.
For $M_0$ non-zero eigenvalues, the energies are determined from $E /\Delta=1 /\sqrt{1 + \tan^2 \chi}$ and
\begin{eqnarray}
\label{eq:Reg_tan}
\tan \chi &=& \frac{1}{2} \left\{ \tan (\varphi_{03}/2) + \tan (\varphi_{10}/2) \right\}
\nonumber \\
&&\hspace{-4mm} \pm \frac{1}{2} \sqrt{ \left\{ \tan (\varphi_{03}/2) - \tan (\varphi_{10}/2) \right\}^2 + 4 g_i |h|^2}.
\end{eqnarray}
Since this expression is only valid in the vicinity of $\varphi_{\rm c} = 2\pi/7$,
we need to expand in this vicinity in terms of small $\phi = \varphi - \varphi_{\rm c}$ and
take $h$ as a function of $\phi$. This gives
\begin{equation}
E_{g_i, \pm} = E_{\rm c} + \frac{(C_0+C_3)}{2} \phi \pm \sqrt{g_i|h (\phi )|^2 + \frac{(C_0-C_3)^2}{4} \phi^2},
\label{eq:Reg_E1}
\end{equation}
where coefficient $C_{0,3}$ and $|h|$ are of the order of $1$, their concrete values are of no interest now.
This makes the quasi-hyperbolic phase dependence of the energies and the absence of
irregular fluctuations explicit. In this form, the expression describes the vicinity of any crossing point of regular type.

\begin{figure}[t] 
\begin{center}
\includegraphics[width = \columnwidth,angle=0]{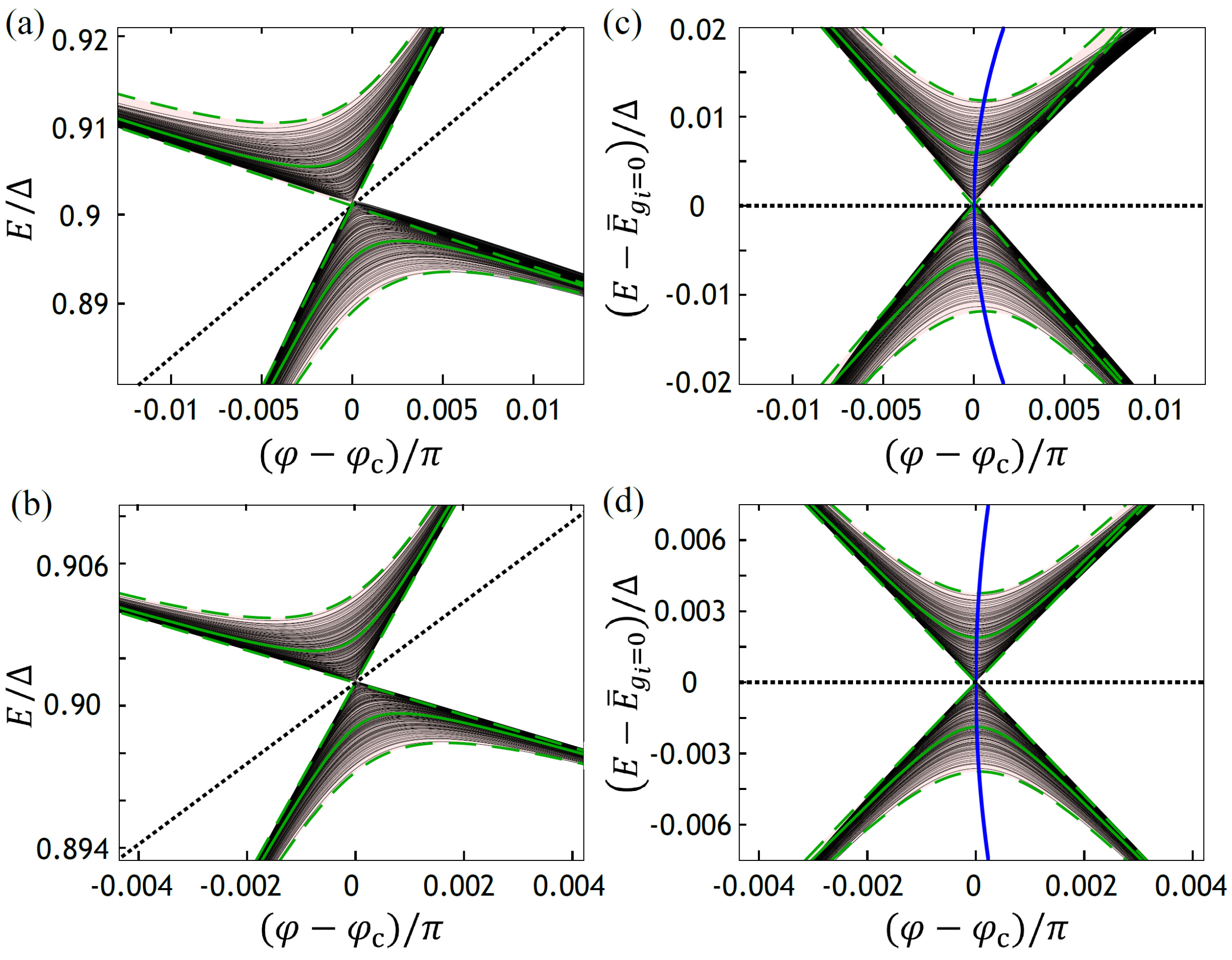}
\end{center}
\caption{(Color online)
Comparison of the first-order perturbation results in Eq.\ (\ref{eq:Reg_E1}) with
full numerical results for two ratios $r=10^{-3}$ [panels (a) and (c)] and $r=10^{-4}$ [panels (b) and (d)]
in the vicinity of one regular crossing at $\varphi = 2\pi /7 \equiv \varphi_{\rm c}$.
Panels (a) and (b) demonstrate the Andreev levels (thin black curves) and
analytical results by the perturbation with $g = \langle g_i \rangle = r$ (thick solid green line) and
$g = 0, 4r$ (thick dashed green line).
The dotted line in panels (a) and (b) indicates the half-sum of the two energies for $g_i = 0$,
$\bar{E}_{g_i = 0} = ( E_{g_i = 0 ,+} + E_{g_i = 0 ,-})/2$, namely the first two terms in Eq.\ (\ref{eq:Reg_E1}).
Panels (c) and (d) are corrected results by substracting the half-sum from Eq.\ (\ref{eq:Reg_E1}).
The blue solid curve in (c) and (d) indicates the points of minimum separation of two anticrossing levels for various $g_i$.
This curve implies asymmetric deviation with respect to $\varphi - \varphi_{\rm c}$.
The dotted lines indicate zero.
}
\label{fig:1OrderModel}
\end{figure}

The randomness of the setup is manifested in the randomness of the eigenvalues $g_i$.
In the limit of weak reflection, $M/N \equiv r \ll 1$,
the matrix $G^{(0)\dagger}_{30}$ can be regarded as a member of the Gaussian ensemble~\cite{Campagnano}.
As it has been shown in Ref.\ \cite{Campagnano}, in the limit of big numbers of channels $M_0,M_3 \approx M$
the distribution of eigenvalues has the specific form
\begin{equation} 
\label{eq:distribution}
\rho(g) = \Theta(4r -g)\frac{M}{2\pi r} \sqrt{\frac{4 r}{g}-1}. 
\end{equation}
This distribution is bounded by $4r$, the average $\langle g\rangle =r$ and
equals to the standard deviation $\sqrt{\langle g^2\rangle-\langle g\rangle^2}$.
The eigenvalue density diverges at $g \to 0$ and vanishes upon approaching $4r$.

Upon the end of this subsection, let us compare in detail the above analytical results with numerical ones.
Figure \ref{fig:1OrderModel} demonstrates the regular crossing with the comparison for two different ratios
$r = 10^{-3}$ [(a) and (c)] and $r = 10^{-4}$ [(b) and (d)] in the vicinity of a crossing at $\varphi_{\rm c}=2\pi /7$.
We assume that the numbers of channels in all nodes are equal, $M_i = 100$.
Thin black curves in Figs.\ \ref{fig:1OrderModel}(a) and (b) give the ABS energies by numerical calculation for
a single realization of the random scattering matrices.
The analytical results by up to the second order perturbation explain the regular crossing behavior for
an interval of eigenvalue distribution of the paraeter $g_i$, $[0, 4r]$.
We find a small deviation of the Andreev levels from the interval region above a line $E_{g_i =0, -} (\phi)$,
especially at $\phi >0$ in Fig.\ (a). This deviation is strongly suppressed for smaller $r$ in (b).
Let us replot the levels by substracting a linearly increasing component with $\phi$,
$\bar{E}_{g_i = 0} = ( E_{g_i = 0 ,+} + E_{g_i = 0 ,-})/2 = E_{\rm c} + (1/2) (C_0 + C_3) \phi$, from
both the analytical and numerical ones.
In Figs.\ \ref{fig:1OrderModel}(c) and (d), the spectrum looks quasi-hyperbolic with respect to $\phi$.
However, the minimal point of difference between the lower and upper energies shifts
slightly to positive $\phi$ with the increase of $g_i$ from zero.
This shift is owing to $\phi$-dependence of $|h|^2$ in the third term in Eq.\ (\ref{eq:Reg_E1}).

First order corrections are important for crossings between level bunches formed in adjacent QPC. 
These contributions vanish at crossings between level bunches of non-adjacent QPC.
Also, even for the case of two adjacent QPC, the parameter $h$ can vanish at the crossing point,
as it does for the crossing exemplified in Fig.\ \ref{fig:crossing}(b). 
In all these situations, as well as far from the crossings, the degeneracy is lifted by second-order terms. 

\subsection{Fine structure of a bunch}

\begin{figure}
\includegraphics[width=\columnwidth]{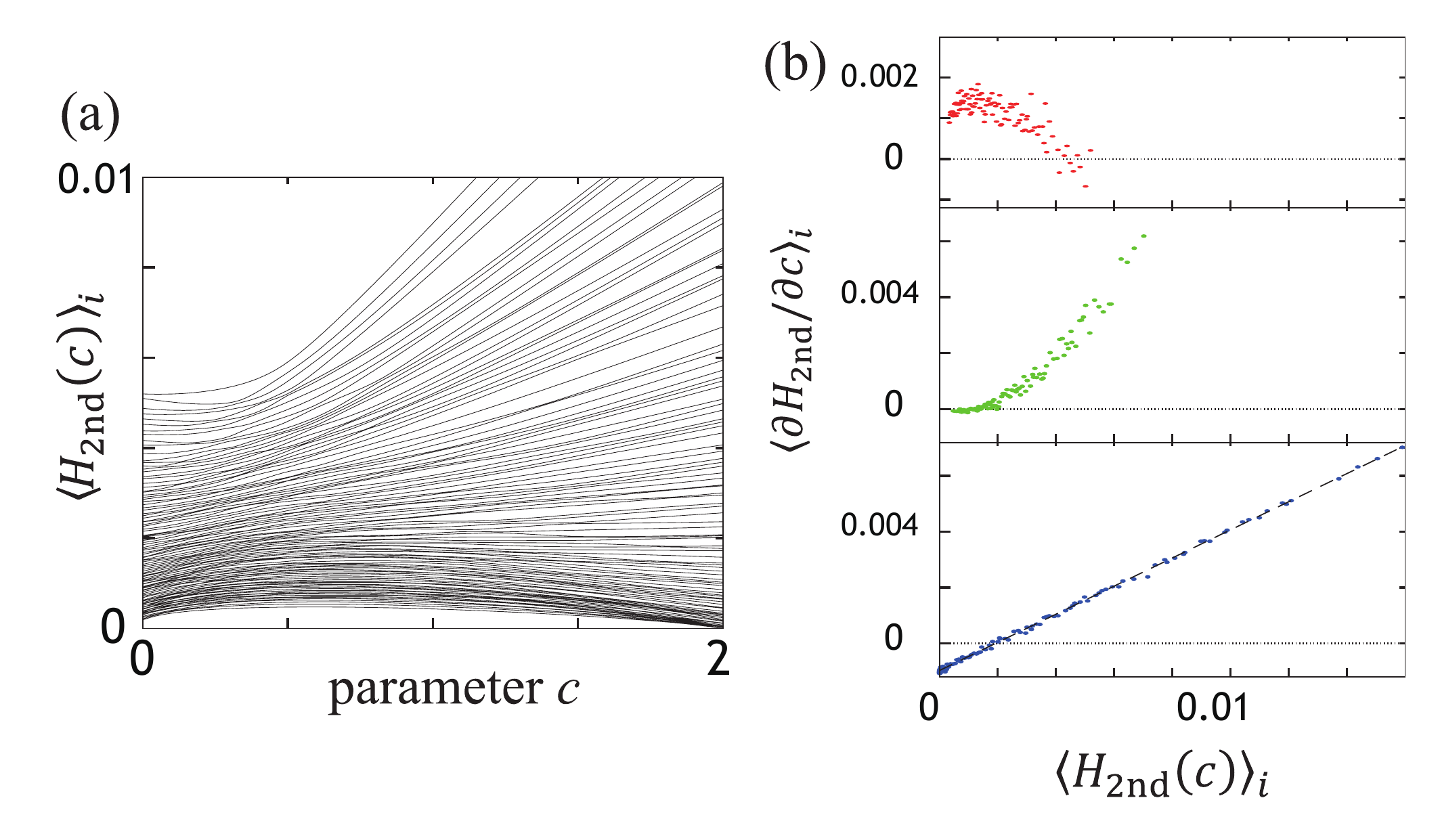}
\caption{
Parametric dependence of the eigenvalues of $H_{\rm 2nd} (c)$ given by Eq.\ (\ref{eq:correction2forc}).
Transmission and reflection matrices in the expression are computed from a random realization
of the scattering matrices. We set $M = 100$ (matrix size is $100 \times 100$).
The ratio is $M/N = 10^{-3}$.
(a) The eigenvalues of $H_{\rm 2nd} (c)$. In $0<c<2$, they are all positive.
(b) The velocities versus the eigenvalues  at $c = 0.1$ (top), $0.7$ (middle),
and $2$ (bottom panel). In the bottom panel, we also give a linear fit.
}
\label{fig:paramerize}
\end{figure}

We start our consideration of the second order corrections with the degeneracy lifting in a bunch far from the crossing points.
As a concrete example we take the bunch of degenerate levels in the QPC 0 with zero-order energies $E/\Delta = \cos (\varphi_{10}/2)$.
The consideration of other bunches is similar. As discussed, the first-order terms vanish.
The second-order terms, given by Eq.\ (\ref{eq:correction2}), are collected into the following matrix:
\begin{eqnarray}
H_{{\rm eff}} &=&
f(\varphi_{10}, \varphi_{03}) t_{31}^{(0)} t_{31}^{(0)\dagger}
  + f(\varphi_{10}, \varphi_{21}) t_{02}^{(1)\dagger} t_{02}^{(1)}  \nonumber \\
&& \hspace{5mm} + f(\varphi_{10}, \varphi_{01})
\left( r_{00}^{(1)} - r_{11}^{(0)\dagger} \right) \left( r_{00}^{(1)\dagger} - r_{11}^{(0)} \right)
\label{eq:correction2for1}
\end{eqnarray}
where the factors are defined as
\begin{equation}
f(a,b) \equiv \frac{\sin (a/2) \sin (b/2)}{\cos^2 (a/2) \sin (a/2 - b/2)}.
\end{equation}
The transmission and reflection matrices in Eq.\ (\ref{eq:correction2for1}) are
components of the scattering matrices $s^{(0)}$ and $s^{(1)}$.
For instance, $t_{31}^{(0)}$ describes the scattering of  an electron moving from node $1$ to $3$ via node $0$.
The split energie levels in the bunch are directly related to the eigenvalues of $H_{{\rm eff}}$.

\begin{figure}
\includegraphics[width=0.9\columnwidth]{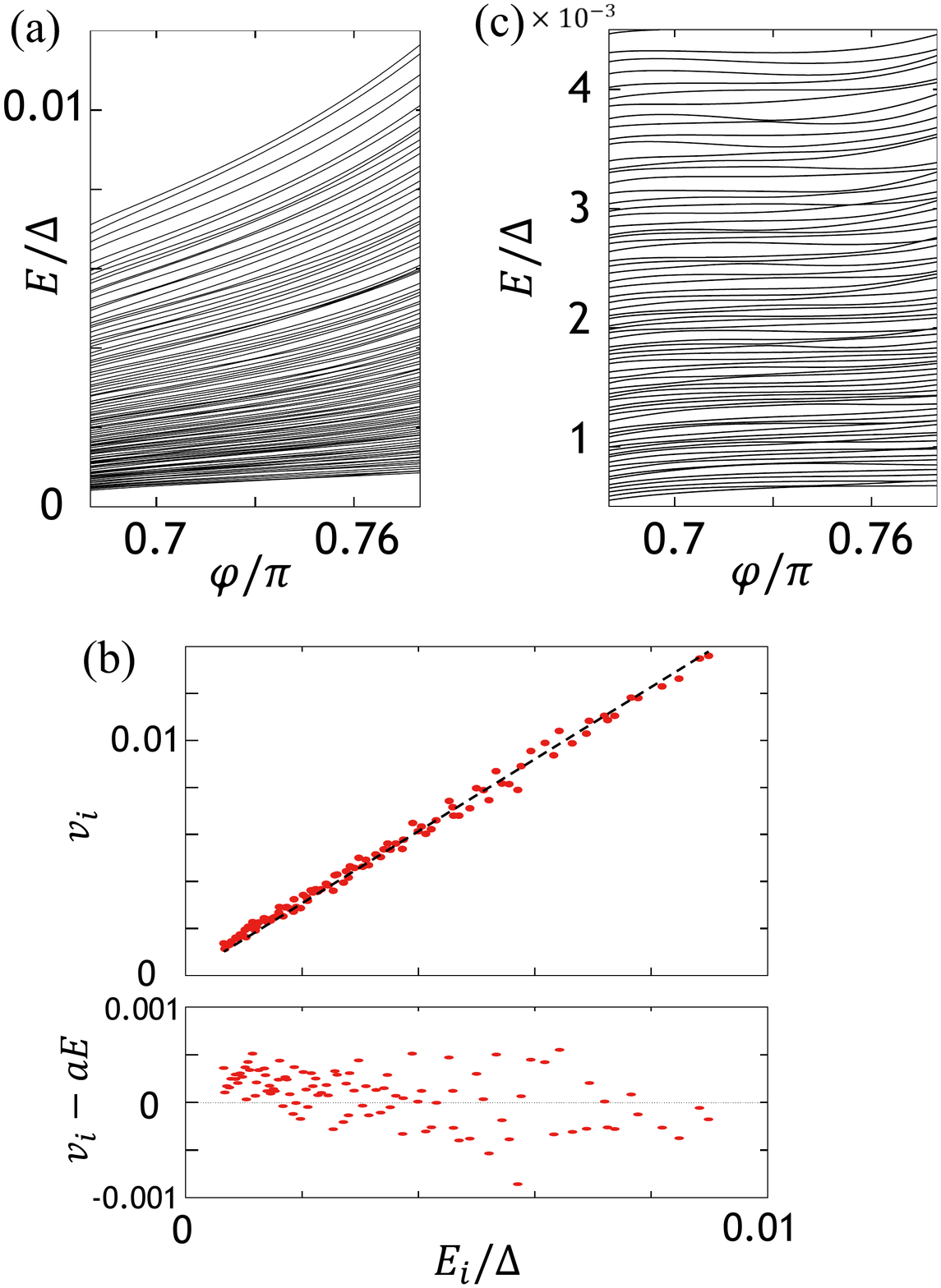}
\caption{
Fine structure of the bunch ($M=100$, $M/N=10^3$).
(a) The  ABS energy shifts from $\cos (\varphi_{10}/2)$ in a narrow interval centered at
$\varphi = 0.73 \pi$ [see Fig.\ \ref{fig:ABS1}(a) for a bigger plot].
(b) The level velocities $v_i = \langle (ie^{-i2\chi} \sqrt{1-(E/\Delta)^2}/2) \partial S/\partial \varphi \rangle_i$ at
$\varphi = 0.73\pi$ versus the energy shifts $E_i$.
The fitting line is given by $v = a (E/\Delta)$ ($a \simeq 1.53412$).
The lower panel shows the velocities upon subtracting the linear fit.
(c) ABS energies upon subtracting $E_i(\varphi = 0.73\pi) \{ \exp (a (\varphi - 0.73\pi)) -1\}$ from each curve.
The remaining dependence is mostly irregular.
}
\label{fig:isolated}
\end{figure}

$H_{{\rm eff}}$ is a linear superposition of three random (positively defined) matrices that
do not depend on phases, the factors $f$ are smooth functions of phases.
One of the $f$s diverges upon approaching a regular crossing indicating
the increasing importance of first-order corrections. 
Since the matrices are random, and generally do not commute,
we expect to see the irregular dependence on the phase imposed on a regular one.
The typical eigenvalues of the matrices are $\simeq r \equiv M/N$.
Since $f \simeq 1$, the regular level velocities are roughly estimated as $v_r \simeq r$,
and irregular ones as $v_{ir} \simeq r/\sqrt{M}$.
However, there are reasons to doubt this estimation, since it is not evident how non-commutative the matrices are.
Besides, the relative degree of irregularity should depend on relative magnitudes of different $f$:
if one of the factors dominates, the eigenvalues of $H_{{\rm eff}}$ are determined by
the eigenvalues of a single matrix and are therefore smooth functions of phase.

To comprehend this with an example, we first consider a modification of
Eq.\ (\ref{eq:correction2for1}) that depends on a single parameter $c$;
\begin{eqnarray}
H_{\rm 2nd} (c) &=& (1-c/2) t_{31}^{(0)} t_{31}^{(0)\dagger}
+ (1-c/2)t_{02}^{(1)\dagger} t_{02}^{(1)} \nonumber \\
&& \hspace{5mm}
+ c \left( r_{00}^{(1)} - r_{11}^{(0)\dagger} \right)
\left( r_{00}^{(1)\dagger} - r_{11}^{(0)} \right).
\label{eq:correction2forc}
\end{eqnarray}
Figure \ref{fig:paramerize} (a) shows the dependence of the eigenvalues $\langle H_{\rm 2nd} \rangle_i$ on $c$.
Since $H_{\rm 2nd}$ is positively defined for $0 \leq c \leq 2$, all eigenvalues are positive in this interval.
The spectrum is more dense at smaller eigenvalues.
For small $c$, one can see the irregular dependence in the form of level wiggles.
For $c>1$, the third term in $H_{\rm 2nd}$ tends to dominate.
In this case, the spectrum shows a quasi-linear regular dependence of the eigenvalues. 
At $c = 2$, the first and second terms vanish, $H_{\rm 2nd}$ is given by a single positively defined matrix,
and the lower boundary of the distribution is close to zero.
To quantify this, we plot in Fig.\ \ref{fig:paramerize}(b) the level velocities
$\langle \partial H_{\rm 2nd}/\partial c \rangle_i$ versus the corresponding eigenvalue.
The dots are randomly distributed around a smooth curve.
For the calculation we took $M=100$ and $r=10^{-3}$, so the above rough estimation gives $v_r \simeq 10^{-3}$,
$v_{ir}/v_r \simeq 0.1$. This is qualitatively valid at small values of the parameter $c$. 
At larger $c$, the irregular part of the velocities is hardly visible at the regular background.
We thus conclude that the domination of one of the terms in the Hamiltonian efficiently quenches
the irregular dependence of the eigenvalues.

Let us now turn to the dependence of the energies on the real phases rather than on the factors.
In the interval between $\varphi = 2\pi /3$ and $4\pi /5$, all factors in Eq.\ (\ref{eq:correction2for1}) are negative,
so that $H_{{\rm eff}}$ in Eq.\ (\ref{eq:correction2for1}) is a negatively defined matrix.
This gives positive shifts of the ABS energies $E_i$ with respect to the zero-order value $\Delta \cos (\varphi_{10}/2)$.
We plot the shifts in Figure \ref{fig:isolated}(a). We concentrate on a narrow interval around $\varphi =0.73 \pi$,
where it is easier to distinguish the regular and irregular dependences of the energies and
where the rough estimation predicts about one wiggle per level.
Even in this relatively small interval, the band width of the ABS energies in Fig.\ \ref{fig:isolated}(a) changes significantly.
The visible phase dependence is mostly regular. In Fig.\ \ref{fig:isolated}(b),
we plot the velocities of the ABS levels at a fixed phase $\varphi = 0.73\pi$ versus the energy shifts.
We see that apart from its irregular component, the velocity is approximately proportional to
the corresponding energy shift. We understand from the previous example that
this signifies the dominance of one of the three terms in $H_{{\rm eff}}$.
In the lower panel of Fig.\ \ref{fig:isolated}(b), we substract the linear fit revealing the irregular part.
Its magnitude conforms the estimations. 
The regular part of the phase dependence can be fitted by an exponential function of $\varphi$.
We subtract the estimated exponential function, $\sim \exp ((a (\varphi - 0.73\pi ))$, from
the ABS energy shifts in Fig.\ \ref{fig:isolated}(c) revealing the irregular dependence on the phase.
This dependence looks like a standard expectation for a parametric dependence derived from the RMT~\cite{Simons}. 

\subsection{Irregular crossings}

We extend the discussion of the second-order corrections to crossing points.
As an example, we concentrate on the crossing point presented in Fig.\ \ref{fig:crossing}(b).
The bunches from QPC 0 and QPC 1 cross here.
One would expect first-order terms coming from the scattering between these channels via the common node 1.
However, at the particular line in phase space the coefficient $h(\varphi_{01},\varphi_{12})$ in front of
these terms (c.f. Eqs.\ (\ref{eq:Reg_H}) and (\ref{eq:H})) vanishes at the crossing point.
We need to investigate the second-order terms. 
They may be arranged in a block structure corresponding to states in QPC 0 and 1,
\begin{equation}
H_{{\rm eff}} =\left( \begin{array}{cc}
H_{00} & H^\dagger_{10} \\
H_{10} & H_{11}
\end{array} \right)
\end{equation}

We have already evaluated one of the diagonal blocks in the previous subsection,
$H_{00} \equiv H_{{\rm eff}}$ as given by Eq.\ (\ref{eq:correction2for1}).
Another block is obtained by the index permutation:
\begin{eqnarray}
H_{11} &=&
f(\varphi_{12}, \varphi_{01}) t_{20}^{(1)\dagger} t_{20}^{(1)}
  + f(\varphi_{12}, \varphi_{23}) t_{31}^{(2)} t_{31}^{(2)\dagger} \nonumber \\
&& \hspace{5mm} + f(\varphi_{12}, \varphi_{21})
\left( r_{22}^{(1)} - r_{11}^{(2)\dagger} \right) \left( r_{22}^{(1)\dagger} - r_{11}^{(2)} \right),
\end{eqnarray}
while the non-diagonal block is given by 
\begin{eqnarray}
H_{10} &=& g(\varphi_{12}, \varphi_{10}) \left( r_{22}^{(1)} - r_{11}^{(2)\dagger} \right) t_{02}^{(1)}
\nonumber \\
&& \hspace{5mm}
+ g(\varphi_{10}, \varphi_{12})^* t_{20}^{(1)\dagger} \left( r_{00}^{(1)\dagger} - r_{11}^{(0)} \right),
\end{eqnarray}
where
\begin{equation}
g(a,b) \equiv \frac{\sin (a/2)}{ 2\cos^2 (a/2) \cos (b/2)} e^{i (a-b)/2}.
\end{equation}

\begin{figure}
\includegraphics[width=0.9\columnwidth]{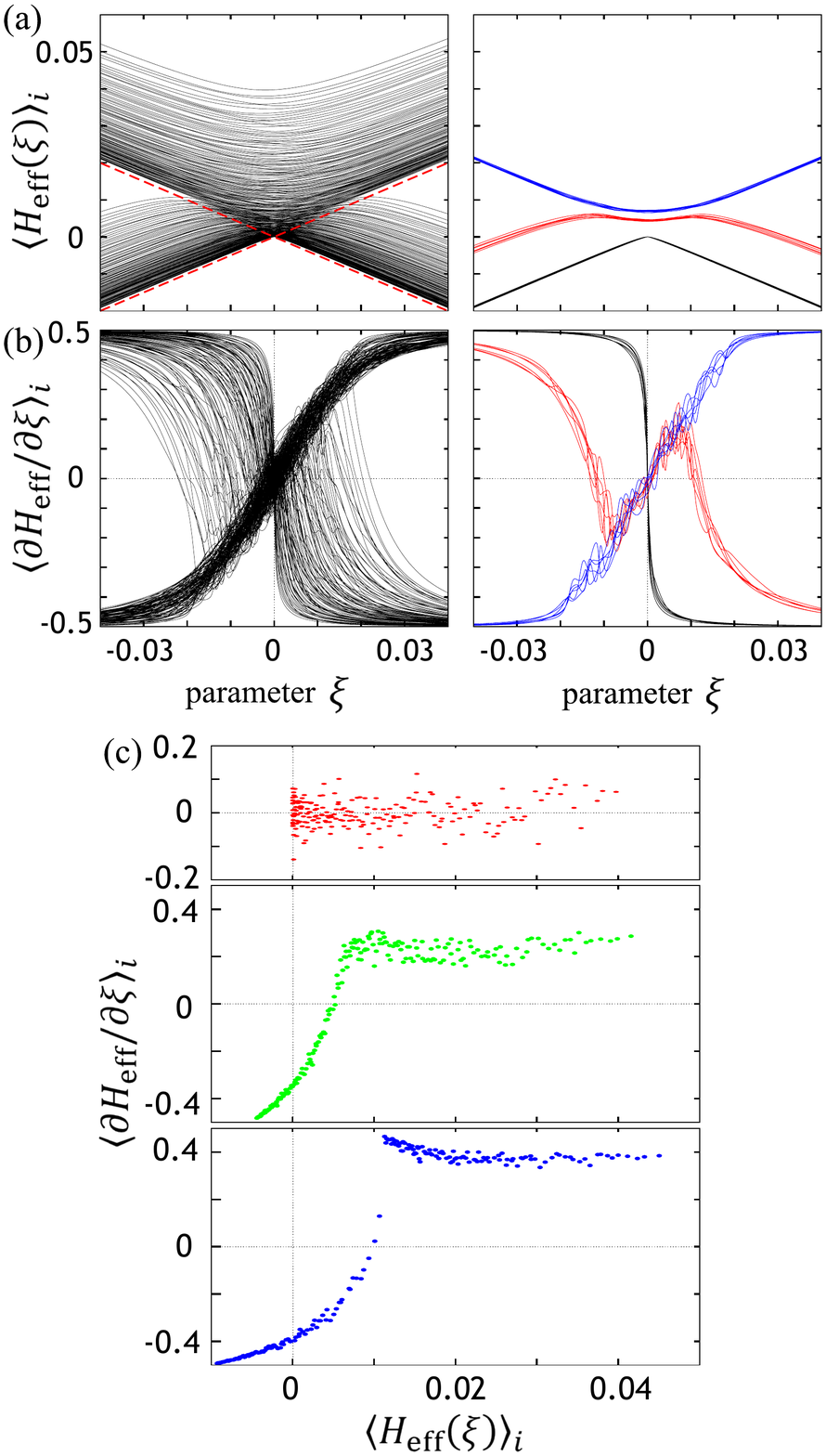}
\caption{
Irregular crossing. We plot the eigenvalues and the velocities of $H_{\rm eff} (\xi )$ given by
Eq.\ (\ref{eq:simpleH}). The parameters are the same as in Fig.\ \ref{fig:paramerize}.
The matrix size of $H_{\rm eff} (\xi )$ is $200 \times 200$ ($M_0 = M_1 = 100$).
(a) Eigenvalues of $H_{\rm eff} (\xi )$. The left panel shows all eigenvalues.
Dashed red lines in the left panel correspond to $H=\pm \xi /2$.
In the right panel, we plot three groups of selected eigenvalues with indexes
$i = 1\sim 6$ (black), $81 \sim 86$ (red),
and $101 \sim 106$ (blue).
(b) Velocities of the corresponding eigenvalues (a). 
(c) The velocites versus the eigenvalues  $\xi = 0$ (top),
$0.01$ (middle), and $0.02$ (bottom panel).
}
\label{fig:cross}
\end{figure}

We need to evaluate this matrix at the crossing point $\varphi_{\rm c} = 2\pi /3$,
where  $e^{i\varphi_{10}} = e^{i\varphi_{12}}$ and
$f$ and $g$ thus satisfy
\begin{equation}
f(\varphi_{10},\varphi_{21}) f(\varphi_{12},\varphi_{21}) = | g(\varphi_{12},\varphi_{10}) |^2.
\end{equation}
With this, the matrix can be presented in the form
\begin{equation}
H_{{\rm eff}} = A A^\dagger + B B^\dagger
\label{eq:positiveQ}
\end{equation}
with
\begin{eqnarray}
A &\equiv & \left( \begin{array}{cc}
\sqrt{ f_{10,03} } t_{31}^{(0)} & \\
     & \sqrt{ f_{12,23} } t_{31}^{(2)\dagger}
\end{array} \right), \\
B &\equiv & \left( \begin{array}{cc}
e^{i\alpha} \sqrt{ f_{10,21} } t_{02}^{(1)\dagger}
    & e^{-i\beta} \sqrt{ f_{10,01} } \left( r_{00}^{(1)} - r_{11}^{(0)\dagger} \right) \\
e^{i\beta} \sqrt{ f_{12,21} } \left( r_{22}^{(1)} - r_{11}^{(2)\dagger} \right)
    & e^{-i\alpha} \sqrt{ f_{12,01} } t_{20}^{(1)\dagger}
\end{array} \right).
\end{eqnarray}
Here we introduced the abbreviation $f_{ij,kl} \equiv f(\varphi_{ij},\varphi_{kl})$.
The representation (\ref{eq:positiveQ}) makes explicit that the second-order matrix is positively defined.

To consider the vicinity of the crossing point, we add the zero-order terms.
It is convenient to incorporate these terms into the parameter $\xi$ proportional to
the phase deviation from the crossing point so that the resulting matrix reads 
\begin{equation}
H_{\rm eff}(\xi) = \frac{1}{2}
\left( \begin{array}{cc}
\xi (\varphi ) &  \\
                   & -\xi (\varphi )
\end{array} \right)
+H_{\rm eff}.
\label{eq:simpleH}
\end{equation}
In Fig.\ \ref{fig:cross}, we present the eigenvalues [Fig.\ \ref{fig:cross}(a)]
and the eigenvalue velocities [Fig.\ \ref{fig:cross}(b)] of $H_{\rm eff}(\xi)$.
Since $H_{\rm eff}$ is positively defined, $-|\xi|/2$ is the precise lower boundary of the resulting spectrum,
this is clearly seen in the spectrum. Far from the crossing, the levels are separated into two bands.
The eigenvalues are readily given by $\xi/2 + H^i_{00}$, $-\xi/2 + H^i_{11}$, $H^i_{00,11}$ being
the eigenvalues of the two diagonal blocks.
The $\xi$-dependence is thus very regular far from the crossing.
The eigenvalues of the diagonal blocks are distributed according to Eq. (\ref{eq:distribution}).
This explains the rather definite width of the bands and the concentration of the eigenvalues at the lower edges.

The $\xi$-dependence is clearly irregular directly at the crossing where two bands merge,
while the degree of the irregularity depends on the position of the level with respect to
the lower boundary of the spectrum.
We illustrate the latter in the right panels of Fig.\ \ref{fig:cross}(a) and (b) selecting three groups of few levels.
The lowest 6 levels closely follow  the lower boundary $-|\xi |/2$,
their velocities changing sharply at $\xi \approx 0$ exhibiting no visible irregularities.
For the group of the 6 levels close to the upper egde of the lower band,
the regular part of the velocity changes non-monotonically changing sign near the openings of the smile gaps,
while the velocity of the group closer to the lower edge is monotonic.
The groups come close to each other in the interval $-0.01<\xi<0.01$ where the bands merge.
The irregular dependence with about a dozen wiggles is observed in a twice bigger interval.

Figure \ref{fig:cross}(c) demonstrates a correlation between the eigenvalues and their velocities.
At $\xi = 0$ (top panel), the regular part of the velocity cancels owing to symmetry and substantial irregular variations are seen.

Close to the opening of the smile gap, at $\xi = 0.01$ (middle panel),
the lower levels have negative velocities, starting with $-1/2$.
The velocities increase upon  increasing the eigenvalue saturating at $\approx 0.25$
where they still exhibit significant irregularities. 
For $\xi =0.02$ where the bands already separated we see the velocities reaching $1/2$
at the lower edge of the upper band. 
The velocity distribution is clearly divided into two groups corresponding to the bands.

\section{Conclusions}
\label{sec:conclusions}

We have proposed a setup of a multi-terminal superconducting nano-device, {\it 4T-ring},
that has an interesting and complex spectrum of ABS and exemplifies 
the opportunities of nano-design in such structures.
The spectrum can be readily tuned by the superconducting phases of the terminals and is defined in a 3D parametric space of independent phases.
The properties of the spectrum crucially depend on
the ratio between conductances (or numbers of channels) of the quantum point contacts inside the ring and those connected to the superconducting terminals, 
$G^{\rm i}/G^{\rm o}$.
The spectrum exhibits a variety of gaps:
the proximity gaps that open at zero energy and smile gaps where the levels are present below and above the gap.
We have investigated in detail the spectrum demonstrating gaps with an irregular parametric dependence of
the ABS energies in combination with rather ordered gaps.
While disorder-specific manifestations are typical for a generic random system, the order emerges from the rich topological properties that are specific for the setup.
The topological nature of the system protects the existence of proximity and smile gaps.

The topology of the first kind is associated with the proximity gaps.
The semiclassical Greens function at zero energy, which gives the density of states, is associated with
5 topological numbers, $n_4$ and $n_{0,1,2,3}$.
They satisfy the relation $n_4 = \sum_{i = 0,1,2,3} n_i$ suggesting a similarity with
the classification of topological insulators.
The number $n_4$ and the set of $n_i$ distinguish the gapped and gapless regions in
the 3D parameter space of phases $\varphi_i$.
At small values of the ratio (open limit), the proximity gap is open almost everywhere except the vicinities of the special points $\varphi_{ij} = \pi$.
The gaped regions separated by the points can be labeled by the 4 topological numbers $n_{0,1,2,3}$.
At big values of the ratio (closed limit), several regions become gapless, where $n_4 \ne 0$.

The topology of the second kind emerges from the fact that the transmission distributions of
the device nodes have gaps at low transmissions in the open limit.
This protects four topological numbers that are numbers of transport channels in each inner QPC in the device.
We explain the existence and properties of the smile gaps by making use of those rather concealed topological numbers.
The smile gap can be punctured by injecting artificial transmission eigenvalues in the gap of the transmission distribution. This opens up unique design opportunities to generate isolated levels in a quasi-continuous spectrum.

The spectrum is highly degenerate in the extreme open limit of very conductive outer QPCs.
The ABS levels are grouped in narrow bunches that cross.
We have discussed a perturbation theory for degenerate levels and
investigated the complex lifting of this degeneracy with an increase of the ratio.
In the vicinity of the crossings, the spectrum demonstrates either regular or irregular behavior depending on the presence of first-order terms.
Random fluctuations of the level spacings upon changing the phase, which is a signature of a random system,
can be seen in isolated bunches as well as at the irregular crossing points.

The proposed system realizes new kinds of topology in mesoscopic physics, to be compared with,
for instance, Majorana fermions~\cite{Mourik} and Weyl singularities~\cite{riwar:16,YokoyamaNazarov2015}.
The topologies in the 4T-ring do not require the use of exotic materials
and illustrate the potential of multi-terminal superconducting structures.

\section*{ACKNOWLEDGMENT}
We appriciate the fruitful discussion with Roman-Pascal Riwar, Manuel Houzet, Julia S.\ Meyer, Leonid Glazman.
This work has been partially supported by JSPS Postdoctoral Fellowships for Research Abroad and
the Nanosciences Foundation in Grenoble, in the framework of its Chair of Excellence program grand in Grenoble. J.~R. and W.~B. were supported by the Carl Zeiss Foundation.

\end{document}